\documentclass[bibyear]{aa}

\usepackage{graphicx}
\usepackage{txfonts}
\usepackage{natbib}
\bibpunct{(}{)}{;}{a}{}{,} 
\usepackage{booktabs}
\usepackage{multicol}
\usepackage{graphicx}
\usepackage{fancyhdr}
\usepackage{amsmath}	
\usepackage{amssymb}	
\usepackage[inter-unit-product=\cdot]{siunitx}
\usepackage{enumerate}
\usepackage{multirow}
\usepackage{mathtools}
\usepackage{longtable}
\usepackage{tabularx}
\usepackage{xcolor} 
\usepackage{ulem} 
\usepackage{comment}
\usepackage{cuted}
\usepackage{soul}
\usepackage{hyperref}
\hypersetup{colorlinks=true,urlcolor=black, citecolor=blue, linkcolor=black}

\makeatletter
\renewcommand*\aa@pageof{, page \thepage{} of \pageref*{LastPage}}
\makeatother

\newcommand{\orcidicon}[1]{\href{https://orcid.org/#1}{\includegraphics[width=11pt]{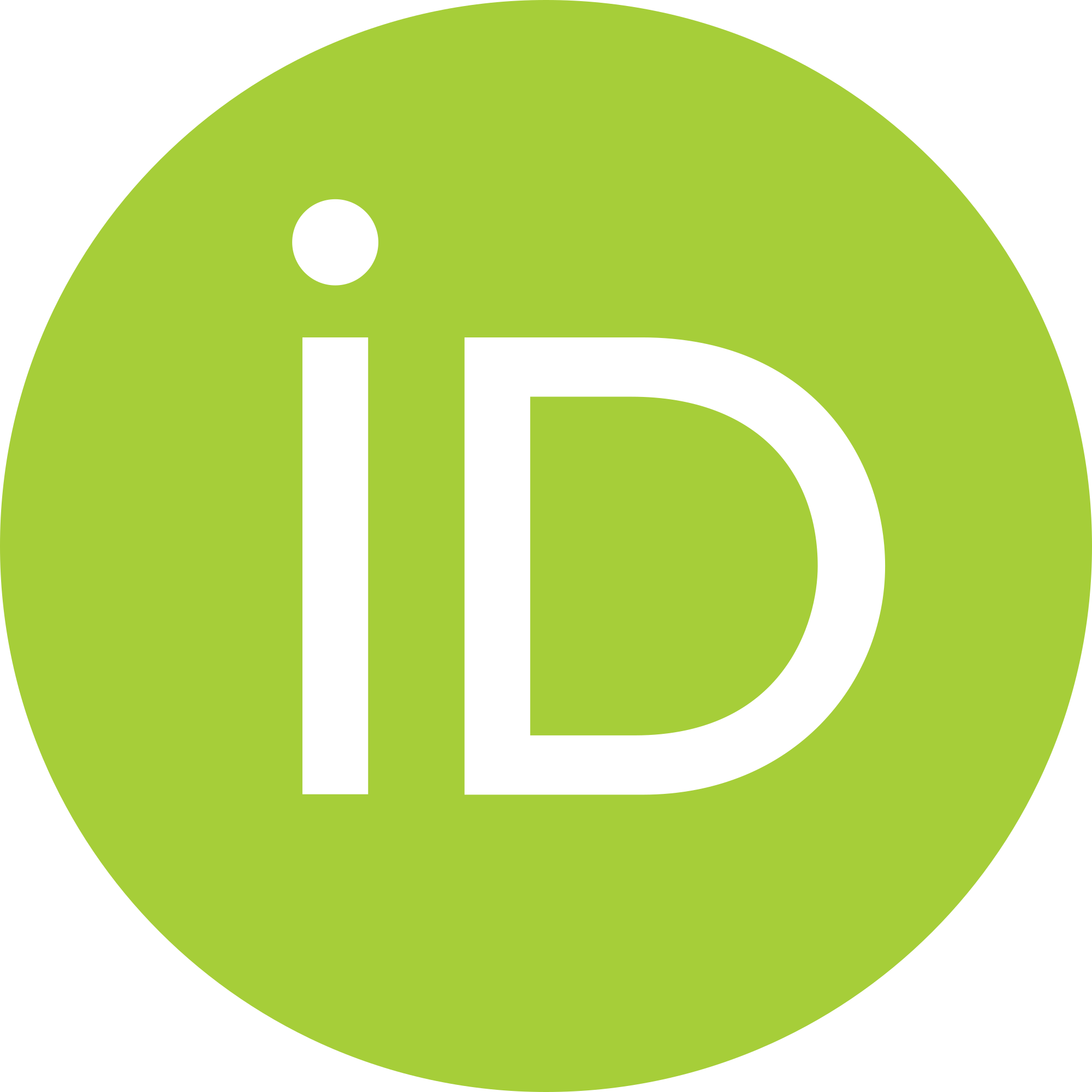}}}
\newcommand{\orcid}[1]{\href{https://orcid.org/#1}{\protect\orcidicon{#1}}}

\newcommand{\fastcluster}{\textsc{fastcluster}}
\newcommand{\clusterbh}{{\sc clusterBH}}
\newcommand{\msun}{{\rm M}_\odot}

\begin{document}

   \title{Hierarchical binary black hole mergers in globular clusters: \\ 
   mass function and evolution with redshift}

    \titlerunning{Hierarchical BBH mergers in GCs}

   \author{
    Stefano Torniamenti\inst{1,2,3,4}
    \orcid{0000-0002-9499-1022} \thanks{\href{mailto:stefano.torniamenti@unipd.it}{stefano.torniamenti@unipd.it}},
    Michela Mapelli\inst{1,2,3,4}
    \orcid{0000-0001-8799-2548}
    \thanks{\href{mailto:mapelli@uni-heidelberg.de}{mapelli@uni-heidelberg.de}},
    Carole Périgois\inst{2,3}, \\
    Manuel Arca Sedda\inst{2,5,6,7},  
    M. Celeste Artale\inst{2,3,8}, 
    Marco Dall'Amico\inst{2,3}, 
    M. Paola Vaccaro\inst{1,2}
    }
    \authorrunning{S. Torniamenti et al.}
    \institute{
    $^{1}$Institut f{\"u}r Theoretische Astrophysik, ZAH, Universit{\"a}t Heidelberg, Albert-Ueberle-Stra{\ss}e 2, D-69120, Heidelberg, Germany\\
    $^{2}$Physics and Astronomy Department Galileo Galilei, University of Padova, Vicolo dell'Osservatorio 3, I--35122, Padova, Italy\\
    $^{3}$INFN - Padova, Via Marzolo 8, I--35131 Padova, Italy\\
$^{4}$INAF - Osservatorio Astronomico di Padova, Vicolo dell'Osservatorio 5, I-35122 Padova, Italy\\
$^{5}$Gran Sasso Science Institute (GSSI), 67100 L’Aquila, Italy\\
$^{6}$INFN, Laboratori Nazionali del Gran Sasso, I-67100 Assergi, Italy\\
$^{7}$INAF - Osservatorio Astronomico d’Abruzzo, Teramo, Italy \\
$^8$Instituto de Astrofisica, Facultad de Ciencias Exactas, Universidad Andres Bello, Fernandez Concha 700, Santiago, Chile
    }
   \date{Received XXXX; accepted YYYY}

 
\abstract{Hierarchical black hole (BH)  mergers are one of the most straightforward mechanisms to produce BHs inside and above the pair-instability mass gap. Here, we investigate the impact of globular cluster (GC) evolution on hierarchical mergers, and we account for the uncertainties related to BH mass pairing functions on the predicted primary BH mass, mass ratio and spin distribution. 
We find that the evolution of the host GC  quenches the hierarchical BH assembly already at the third generation, mainly due to cluster expansion powered by a central BH sub-system. Hierarchical mergers match the primary BH mass distribution from GW events for $m_1 > 50 \, \msun$, regardless of the assumed BH pairing function. 
At lower masses, however, different pairing functions lead to dramatically different predictions on the primary BH mass merger rate density. 
We find that the primary BH mass distribution evolves with redshift, with a larger contribution from mergers with $m_1 \geq 30 \, \msun$ for $z\geq{}2$.
Finally, we calculate the mixing fraction of BBHs from GCs and isolated binary systems. Our predictions are very 
sensitive to the spins, which favor a large fraction ($>0.6$) of BBHs born in GCs, in order to reproduce misaligned spin observations. 
}
  

   \maketitle
%

\section{Introduction}
The first three observing runs of the Advanced LIGO \citep{LIGOdetector} and Virgo \citep{VIRGOdetector} inteferometers have delivered a gravitational-wave (GW) transient catalog (GWTC-3) of $\sim 90$ merger candidates \citep{abbottGWTC1,abbottGWTC2,abbottGWTC-2.1,abbottGWTC3}. 
Most of these events are associated with the merger of binary black holes (BBHs), with masses spanning from the so-called lower mass-gap (e.g., the secondary component of GW190814, \citealp{abbottGW190814}) to the intermediate-mass range, like GW190521 \citep{abbottGW190521,abbottGW190521astro}. The number of detected black holes (BHs) has now become sufficiently large to allow for the exploration of the underlying BH mass distribution \citep{abbottGWTC2population,abbottGWTC3population}, which encodes information about the astrophysical processes governing their formation and merger \citep{fishbach2017,fishbach2021,vitale2017, callister2021,callister2023,farah2023}.

One of the most puzzling features of the primary BH mass distribution from GWTC-3 is the absence of a high-mass cutoff above $\sim60 \, \msun$ \citep{callister2023}, as we would expect from (pulsational) pair instability (\citealp{belczynski2017,woosley2017,spera2017,marchant2019,stevenson2019}). The formation of BHs in this mass range is still matter of debate, because of the large uncertainties that affect both the lower and  upper boundary of the claimed pair-instability mass gap  \citep{farmer2019,farmer2020,mapelli2020,belczynski2020,vanson2020,marchant2020,farrell2020,vink2021,woosley2021,costa2021,briel2023,hendriks2023}. 

In principle, dynamical interactions in star clusters can lead to the formation of BHs in the pair-instability mass gap, through direct star-star collisions \citep{dicarlo2019,renzo2020b,kremer2020,costa2022,ballone2022} or hierarchical (i.e, repeated) BH mergers \citep{quinlan1990,miller2002,oleary2006,antonini2016,fishbach2017,gerosa2017,gerosa2021}. 
Direct $N-$body simulations show that star-star collisions are not rare in young and open clusters \citep{dicarlo2020a,torniamenti2022}, where the short relaxation timescales \citep{portegieszwart2010} lead to mass segregation within less than one Myr, enhancing the possibility of direct star collisions. 
In more massive globular clusters (GCs) and nuclear star clusters, BHs born from the merger of other BHs can  efficiently be retained and pair up again to produce second- and higher- generation mergers. Such hierarchical assembly can repeat several times
and lead to a significant BH mass growth \citep{antonini2016,mapelli2021,mapelli2022,chattopadhyay2023}, thus populating the pair-instability mass gap and intermediate-mass regime.

One of the main challenges to study hierarchical BH mergers in such massive and long-lived clusters is the computational cost. Accurate but computationally expensive Monte Carlo and/or N-body simulations permit to explore only a  limited portion of the parameter space for both the BBH population (e.g., first-generation masses, spins) and the host star cluster (total mass, central density, initial binary fraction). For this reason, fast and flexible semi-analytic models are usually adopted to probe the hierarchical BH assembly \citep{antonini2016,antonini2019,antonini2023,mapelli2021,mapelli2022,kritos2022,kritos2023,arcasedda2023b,fragione2023}. 
In particular, our open-source semi-analytic code \fastcluster{} \citep{mapelli2021,mapelli2022,vaccaro2023}  integrates the evolution of the BBH orbital properties (semi-major axis and eccentricity) driven by GW emission and dynamical encounters, while also giving the possibility to explore different assumptions for the BBH masses, spins, orbital eccentricities, and for different star cluster masses, densities, initial binary fractions. 

In this work, we  account for the temporal evolution of the globular cluster, including  stellar mass loss, two-body relaxation, and tidal stripping by the host galaxy. Our main goal is to explore the impact of GC evolution and  BBH pairing function on the population of hierarchical BBH mergers.  
We investigate how the internal GC evolution and tidal stripping from the host galaxy quench the BH assembly. Also, we account for uncertainties related to first-generation mass distribution by relying on different BBH pairing functions, that is the criterion to pair-up the primary and secondary BH in GCs. 
We investigate the peculiar imprint that hierarchical mergers have on the mass, mass ratio and spin distributions, and we find it can affect the detected parameters from GWs. Finally, we compare the mass, mass ratio and spin distributions of BBH mergers in GCs to those found in isolation, to draw insights on their relative contributions to GW detections.

This work is organized as follows. In Sect. \ref{sec:methods_fc} we introduce the details of the code implementation. Section \ref{sec:results_fc} describes our main results, that we discuss in Sect.~\ref{sec:discussion_fc}. Finally, Sect. \ref{sec:conclusions_fc} summarizes our conclusions.

\section{Methods} \label{sec:methods_fc}

Here, we summarize how we model the evolution of the cluster. We assume the formalism presented by \cite{antonini2020}. 
In particular, we adopt a two-component formalism \citep{breen2013a}, in which the cluster consists of two populations: stars, with mass $m_{\mathrm{\star}}$ and total mass $M_{\mathrm{\star}}$, and BHs, with mass $m_{\mathrm{BH}}$ and total mass $M_{\mathrm{BH}}$. The resulting total star cluster mass is $M_{\rm{tot}} = M_{\mathrm{\star}}+M_{\mathrm{BH}}$.

The total mass of the stellar population decreases because of stellar evolution (stellar winds, supernovae explosions), and tidal stripping from the host galaxy. We model the stellar mass loss as in \cite{antonini2020}, assuming that $M_{\star}$ evolves as a power of time. As a consequence, we assume that the cluster undergoes an adiabatic expansion. 
Also, we implement tidal stripping from the host galaxy following the formalism by \cite{gieles2011a}, who describe the evolution of GCs in circular orbits and in different tidally filling states.

We model  GC expansion due to a central BH sub-system, and relate it to BH ejection from the cluster core \citep{antonini2020}.  
As the GC evolves towards a state of energy equipartition \citep{bianchini2016,spera2016}, BHs segregate to the cluster core, giving birth to a central BH sub-system \citep{spitzer1987}. The temperature difference between BHs and the bulk of the cluster \citep{spitzer1969} generates a heat transfer from the core towards the outer parts \citep{breen2013a}. The  energy flux is produced by BBH hardening, until  dynamical recoils eject most BBHs   \citep{heggie1975,goodman1984}. The GC eventually achieves a state of balanced evolution, where the energy flux from the core is regulated by the energy demand of the cluster as a whole \citep{henon1961}. 

We model the BH mass loss during the phase of balanced evolution, when it can be related to the GC energy transfer \citep{breen2013a}.
We consider the onset of balanced evolution as the time of core collapse of the host GC. Then, we implement the decrease of the total BH mass and the resulting expansion rate following \cite{antonini2020}. More details can be found in 
Appendix \ref{sec:coupling}.

\subsection{Global properties of GCs}\label{sec:clusters}
Following \cite{mapelli2022}, we draw the initial cluster total mass from a log-normal distribution with mean $\langle{}\log_{10}{M_{\rm tot}/{\rm M}_\odot}\rangle{}=5.6$ \citep{harris1996}. We assume a fiducial standard deviation $\sigma_{\rm M}=0.4$. 
We draw the GC density at the half-mass radius from a log-normal distribution with mean $\langle{}\log_{10}{\rho{}/({\rm M}_\odot\,{}{\rm pc}^{-3})}\rangle{}=3.7$, and we assume a fiducial standard deviation $\sigma_\rho=0.4$. In Appendix~\ref{app:M0_rho0}, we will discuss how different initial GC masses and densities affect our results.

We set the core density to $\rho_{\rm c}=20\,{}\rho$ \citep{mapelli2021}. 
This choice yields a good approximation for the ratio of the core to the half-mass density in GCs with the considered BH mass fraction, as shown in \cite{breen2013a}. 
The cluster escape velocity is calculated as \citep{georgiev2009a}:
\begin{equation}\label{eq:vesc}
  v_{\rm esc}=40\,{}{\rm km}\,{}{\rm s}^{-1}\,{}\left(\frac{M_{\rm tot}}{10^5\,{}{\rm M}_\odot}\right)^{1/3}\,{}\left(\frac{\rho}{10^5\,{}{\rm M}_\odot\,{}{\rm pc}^{-3}}\right)^{1/6}.
\end{equation}

\begin{table}
	\begin{center}
	\caption{Summary of the runs.}
	\label{tab:cases}
	\begin{tabular}{lcc} 
	\hline 
    ID & 1g & ng \\ 
    \hline \hline
     A & A & ng1g  \\

     B & B & ngng \\ 
     
     C & B & ng1g \\
  
     \hline
 	\end{tabular}
	\end{center}
	\footnotesize{Column 1: Model name ; column 2: First-generation (1g) mass sampling criterion; column 3: Nth-generation (ng) secondary mass sampling criterion. See Sections~\ref{sec:1gBBH}  and \ref{sec:Ng} for details.} 
\end{table}


\subsection{Hierarchical BBH mergers} \label{sec:fastcluster}
In the following, we present our model for evolving dynamically-formed BBHs in GCs. Since the details of the implementation have already been presented by \citet{mapelli2021}, 
here we mainly focus on the upgrades we introduced.

\begin{figure*}
\begin{center}
\includegraphics[width=\hsize]{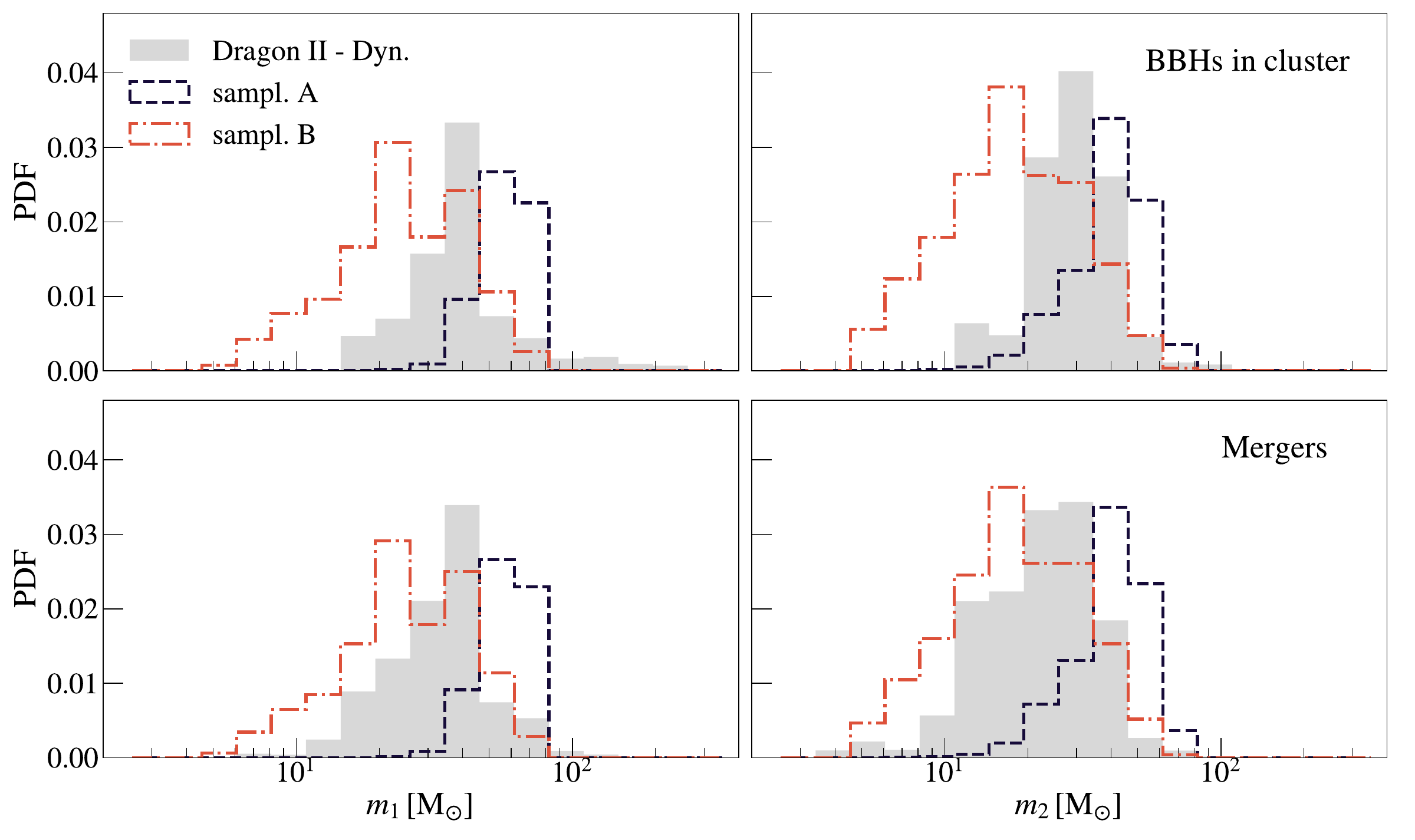}
\caption{Distribution of first-generation primary (left) and secondary (right) masses of dynamical BBHs that are retained within the cluster (upper panels) and merge (lower panels). We show the distributions from the Dragon-II direct $N-$body simulations (\protect\citealt{arcasedda2023d1}, grey shaded area) and in \fastcluster{}, produced by sampling A (blue dashed line) and B (red dot-dashed line).}\label{fig:dragon}
\end{center}
\end{figure*}

\subsubsection{Pairing function and other properties of first-generation (1g) BHs} \label{sec:1gBBH}
We generate the masses of first generation (1g) BHs with the population synthesis code {\sc mobse}\footnote{{\sc mobse} \citep{mapelli2017,giacobbo2018} is a custom and upgraded version of {\sc bse} \citep{hurley2002}, and is publicly available at \url{https://gitlab.com/micmap/mobse_open}.}  \citep{mapelli2017,giacobbo2018}. 
In particular, we adopt the rapid model by \cite{fryer2012} for core-collapse supernovae. For (pulsational) pair-instability supernovae, we use the equations reported in the appendix of \cite{mapelli2020}. 
This yields a minimum BH mass of $\approx 5 \, \msun$ and a maximum BH mass that depends on metallicity, and can be as high as $\approx 70 \, \msun$ for metal-poor stars at $Z=0.0002$, which approximately corresponds to $0.01\,{} {\rm Z}_{\odot}$.

We assume that our 1g BHs pair up with other BHs via three-body encounters at the time of core collapse ($t_{\mathrm{cc}}$, eq.~\ref{eq:tcc}), that is when dynamical interactions are expected to become effective within the BH sub-system core \citep{heggie2003}. 

We pair the BH masses following the mass sampling  criterion from \cite{antonini2023}. This pairing function is conceived to describe the dynamical BBH formation in clusters from three-body encounters, and favours the pairing of the most massive objects. The BH sampling is based on the formation rate of hard binaries per unit of volume and energy \citep{heggie1975}, under the assumption that energy equipartition holds in the BH sub-system: 
\begin{equation} \label{eq:Gamma}
    \Gamma_{\mathrm{3b}}(m_1,m_2,m_3) \propto  \frac{n_1 \,{}n_2 \,{}n_3\,{}m_1^4\,{} m_2^4\,{} m_3^{5/2}}{{(m_1+m_2+m_3)^{1/2}(m_1+m_2)^{1/2}}} \,{}\beta^{9/2},
\end{equation}
where $n_{i}=n(m_i)$ is the number density of BHs with mass $m_i$ (with $i=1,$ 2, and 3), while $\beta{}$ is the inverse of their mean internal energy assuming that they are in energy equipartition.

Since eq. \ref{eq:Gamma} is symmetric with respect to $m_1$ and $m_2$, the probability density function of sampling one of the two is:
\begin{equation} \label{eq:pdf_masses}
p_{1,2}(m_{1,2}) = \int_{m_{\mathrm{low}}}^{m_{\mathrm{up}}} \int_{m_{\mathrm{low}}}^{m_{\mathrm{up}}} dm_{3} \,{}dm_{2,1} \,{}\Gamma_{\rm{3b}}(m_1,m_2,m_3),
\end{equation}
where $m_{\rm{low}}$ and $m_{\rm{up}}$ are the lower and upper limit of the mass distribution. We randomly sample two BH masses per time from the {\sc mobse} catalogue following the probability distribution function in eq. \ref{eq:pdf_masses}. Then, we assign the primary (secondary) mass $m_1$ ($m_2$) to the larger (smaller) sampled mass.
Hereafter, we  refer to this mass sampling criterion as to  model A.

To encompass the uncertainties, we also consider the pairing function introduced by \cite{mapelli2021}, who 
uniformly sample the primary BH mass from the BH mass function (as in the {\sc mobse} catalogue), and draw the secondary BH  mass with the probability distribution function defined by \cite{oleary2016}:
\begin{equation} \label{eq:oleary}
    p(m_2) \propto (m_1 + m_2)^4,
\end{equation}
in the interval $[m_{\rm{min}}, m_1)$ where $m_{\rm{min}}=5 \, \mathrm{M_{\odot}}$. Hereafter, we refer to this mass sampling criterion as to sampling B.

In Sect. \ref{sec:sampling_test} and Fig.~\ref{fig:dragon}, we   
compare the BBH masses generated by the aforementioned pairing functions 
with those obtained from direct $N-$body models.

For both sampling criteria, we apply a further correction to the BH masses, based on the supernova kick, calculated as \citep{mapelli2021}:
\begin{equation}\label{eq:SNkick}
    v_{\rm{SN}} = v_{\rm{H05}} \frac{\langle m_{\rm{NS}} \rangle}{m_{\rm{BH}}},
\end{equation}
where $m_{\rm{BH}}$ is the BH mass, $\langle m_{\rm{NS}} \rangle = 1.33 \, \rm{M_{\odot}}$ is the average neutron star mass \citep{oezel2016}, and $v_{\rm{H05}}$ is randomly drawn from a Maxwellian distribution with root-mean square $265 \, \rm{km \,{}s^{-1}}$. If one member of the BBH receives a kick larger than $v_{\rm esc}$ (eq. \ref{eq:vesc}), we assume that it is ejected from the cluster, and we do not integrate its dynamics any further.   

We sample the dimensionless spin magnitudes ($\chi_1$ and $\chi_2$) of the BHs from a Maxwellian distribution with root-mean square $\sigma_{\chi}=0.1$. We choose this distribution as a toy model, because there is no consensus on the spin distribution from first principles \citep[see, e.g.,][]{fuller2019,belczynski2020,bavera2020,perigois2023}. Our choice qualitatively matches the spin distribution inferred by the LIGO--Virgo--KAGRA (LVK) collaboration \citep{abbottGWTC3population}.  
We draw spin directions isotropic over the sphere, because dynamical interactions reset any spin alignments with the orbital angular momentum of the binary system. 

Finally, we sample the initial orbital eccentricity from a thermal probability distribution \citep{heggie1975}:
\begin{equation}
p(e) = 2\,{}e   \quad e \in \rm{[0,1)},    
\end{equation}
and the semi-major axis from:
\begin{equation}
p(a) \propto a^{-1}   \quad a \in \rm{[1,10^{3}]} \, \rm{R_{\odot}}. \end{equation}
Since soft binaries tend to be disrupted by dynamical interactions \citep{heggie1975}, we first check if the newly-generated binary is hard. If it is not, we do not integrate it any further, because we assume it is disrupted by further interactions \citep{mapelli2021}.


\begin{figure*}
\begin{center}
\includegraphics[width=\hsize]{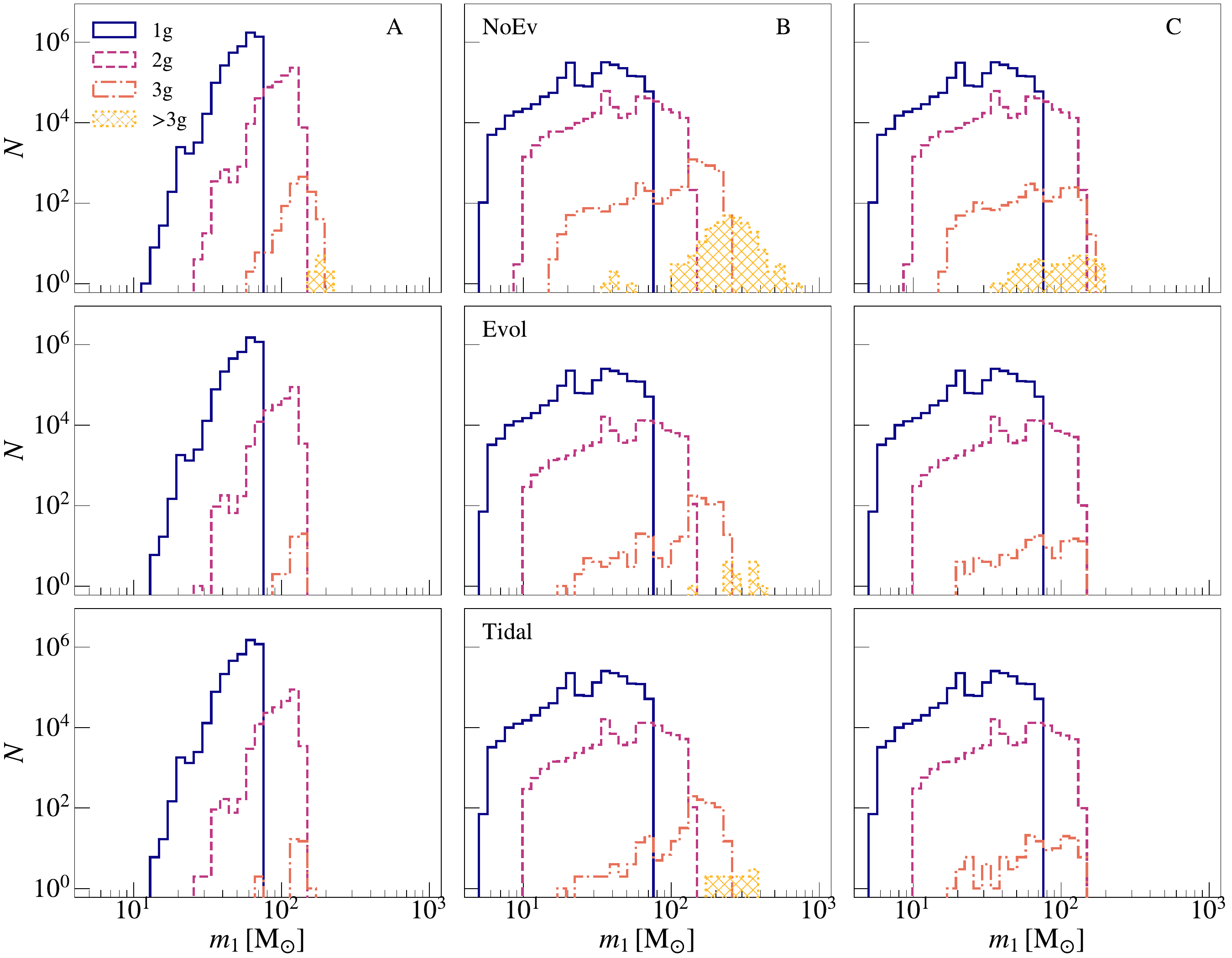}
\caption{Primary BH mass distribution of BBH mergers for models A (left), B (center), and C (right) models, for the three evolutionary cases considered: NoEv (upper panels), Evol (middel panels), Tidal (lower panels).  Different colours correspond to different BH generations: 1g (blue), 2g (purple), 3g (salmon), and $>3$g (yellow) BHs. Here, we show our models with $Z=0.0002$ because they are representative of metal-poor GCs in the Galactic halo \protect\citep{muratov2010}.} 
\label{fig:m1_ng}
\end{center}
\end{figure*}


\subsubsection{Testing the BH pairing functions with \texorpdfstring{$N-$}{N}body models} \label{sec:sampling_test}

In the following, we compare the distributions of BBHs given by the BBH pairing functions introduced in Sect. \ref{sec:1gBBH}, to those from the $N-$body models by  \cite{arcasedda2023d2,arcasedda2023d1,arcasedda2023d3}, whose stellar and binary evolution prescriptions are also based on \cite{hurley2000,hurley2002} formulas (see \citealp{kamlah2022} for more details).
Specifically, we ran $10^6$ models at the same metallicity as the $N-$body simulations ($Z=0.0005$), and compared the distributions of first-generation BBHs that are retained within the cluster and that merge.

As shown in Figure \ref{fig:dragon}, the BBH distributions from $N-$body simulations lie somehow in between the two considered pairing functions. 
The sampling A captures the expected dynamical trend, which favors the coupling of the most massive BHs. The position of the dynamical peak, however, is very sensitive to the underlying stellar evolutionary models. For this reason, different prescriptions can result in discrepancies at the high-mass end of the distribution. 
The $N-$body models also display a long high-mass tail, due to gas accretion from companion stars. 
The sampling B displays a milder increase at higher masses, without the presence of a clear peak, as in the $N-$body models. However, this case can reproduce the trend at low masses, despite the different BH mass distribution at first generation.

Our comparison suggests that there are many uncertainties connected with the BH pairing function, due to the impact of primordial binaries (which are not included in our formalism), deviations with respect to energy equipartition  \citep{spera2016}, stellar and binary evolution models, BH-star interactions, and possibly stochastic fluctuations. 
For example, \cite{arcasedda2023d1} assume that BHs can accrete up to 50\% of stellar mass during BH-star collisions. This results in the high-mass tail above the lower limit of the pair-instability mass gap, which is not present in our models. At the same time, their prescriptions for (pulsational) pair-instability supernovae produce a peak that is $\sim 30\%$ lower than our models at $Z=0.0005$. These differences affect the resulting primary mass distribution, which is expected to peak toward the 1g BH mass upper limit.
At lower masses, different schemes for core collapse supernovae affect the BH mass distribution $\lesssim 30 \, \msun$. In this case, the delayed model used by \cite{arcasedda2023d1} does not predict any mass gap below $5 \, \msun$, in contrast to the rapid model used for our runs. These differences mainly affect the secondary masses, which can extend to the lower limit of the 1g BH mass distribution.

A detailed exploration of all these factors goes beyond the scope of this work and deserves a dedicated study. In the following, we will show the results of both sampling A and B to bracket these uncertainties.

\subsubsection{BBH evolution} \label{sec:bbh_evol}
We evolve the semi-major axis $a$ and orbital eccentricity $e$ of hard BBHs through the formalism described in eq.~16 of   \cite{mapelli2021}, by integrating a system of ordinary differential equations for $a$ and $e$, that takes into account dynamical hardening and GW decay/circularization. 
When a BBH merges within a Hubble time (13.6~Gyr), we estimate the mass and spin of the merger remnant from the fitting formulas by \cite{jimenez-forteza2017}. Also, we calculate the relativistic kick magnitude $v_{\rm K}$  \citep{lousto2012}. The merger remnant is retained inside its parent cluster if the relativistic kick magnitude $v_{\rm K}<v_{\rm esc}$, with $v_{\rm esc}$ given by eq. \ref{eq:vesc}. In this case, we model the dynamical formation of a Nth generation BBH and its evolution, as described in \cite{mapelli2022}.

We iterate the hierarchical merger process until we reach one of the following conditions: the merger remnant is ejected from the cluster, the cluster evaporates, no BHs remain within the cluster, or we reach the Hubble time. 

\subsubsection{Nth-generation secondary mass sampling} \label{sec:Ng}

We sample the Nth-generation (ng) secondary BHs following the same criteria as introduced in Sect. \ref{sec:1gBBH} for the 1g BBHs. 
For the sampling A, we draw $m_2$ from eq. \ref{eq:pdf_masses}, by setting $m_1$ to the mass of the ng primary. Since this model relies on the definition of a probability density function, which cannot be constrained in a direct way for ng BHs, we only consider ng1g BBHs. 

For the sampling B, we draw the secondary mass from eq. \ref{eq:oleary}. We consider two different cases, hereafter {\bf ngng} and {\bf ng1g}. In the {\bf ngng} case, we assume that the secondary BH mass $m_2$ ranges within $[m_{\rm{min}}, m_{\rm 1,ng}]$, where $m_{\rm min}$ is the minimum BH mass and $m_{\rm 1,ng}$ is the mass of the primary BH belonging to the Nth generation. This yields the possibility that the secondary BH comes from previous mergers, as done by \citet{mapelli2021}. 

In the {\bf ng1g} case, we assume that the secondary is always a 1g BH, and $m_2 \in [m_{\rm{min}}, \max{(m_{\rm 1,1g}))}$, where $\max{(m_{\rm 1,1g})}$ is the maximum mass of a primary first-generation BH.
This assumption leads to smaller mass ratios at ng$>$1, because our secondary sampling criterion (eq. \ref{eq:oleary}) favors larger masses but BBHs with $q=1$ can no longer form due to the increasing primary mass. In turn, this can result in larger relativistic recoils at merger, whose magnitude decreases in more symmetric configurations \citep{gonzalez2007,schnittman2007,herrmann2007}.

\begin{figure*}[ht]
\begin{center}
\includegraphics[width=\hsize]{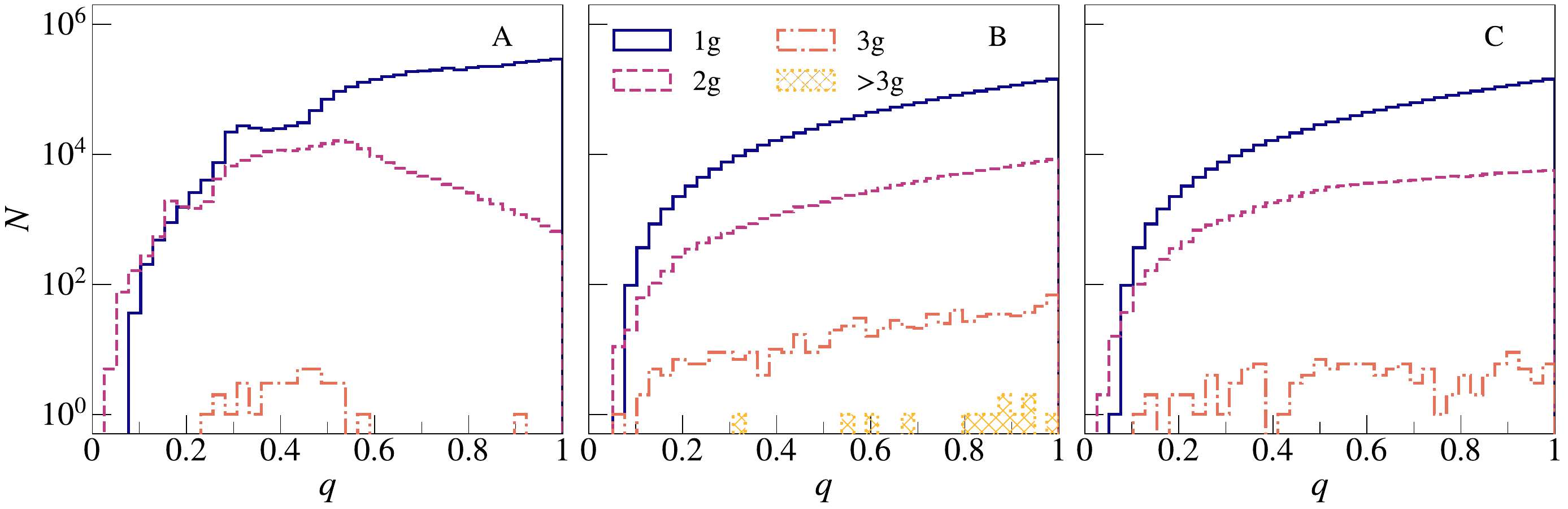}
\caption{Mass ratio distributions of BBH mergers for models A (left), B (center), and C (right) in the Tidal evolutionary case.  Different colours correspond to different BH generations: 1g (blue), 2g (purple), 3g (salmon), and $>3$g (yellow) BHs. Here, we show models with $Z=0.0002$.  
}\label{fig:mass_ratios}
\end{center}
\end{figure*}

\subsection{Merger rate} \label{sec:merger_rate}
We evaluate the merger rate density of dynamical BBHs in GCs  
\citep{mapelli2022}:
\begin{eqnarray}\label{eq:cosmorate}
    \mathcal{R}(z) =\int_{z_{{\rm{max}}}}^{z}
    \,{} \psi{}(z')\,{} \frac{{{\rm d}t(z')}}{{\rm{d}}z'}    \left[\int_{Z_{{\rm{min}}}}^{Z_{{\rm{max}}}} 
    \eta{}(Z)\,{}\mathcal{F}(z',z,Z) \,{}{\rm{d}}Z\right]\,{}{\rm{d}}z',
\end{eqnarray}
where $t(z)$ is the look-back time at redshift $z$, $\psi(z')$ is the GC formation rate density  at redshift $z'$, and $Z_{\rm min}(z')$ ($Z_{\rm max}(z')$) is the minimum (maximum) metallicity of stars formed at redshift $z'$. The merger efficiency $\eta{}$ is defined as the total number of BBHs of a given population that merge within a Hubble time divided by the total initial stellar mass of that population \citep{giacobbo2018,klencki2018}. Finally, $\mathcal{F}(z', z, Z)$ is the merger rate 
of BBHs that form at redshift $z'$ from stars with metallicity $Z$ and merge at redshift $z$, normalized to all BBHs that form from stars with metallicity $Z$. We calculate the cosmological parameters ($H_{0}$, $\Omega_{\rm M}$ and $\Omega_{\Lambda}$) from \cite{planck2016}. 

We assume a Gaussian distribution to model the GC formation rate as a function of redshift:
\begin{equation}\label{eq:GCs}
\psi{}(z)=\mathcal{B}_{\rm GC}\,{}\exp{\left[-\frac{(z-z_{\rm GC})^2}{(2\,{}\sigma_{\rm GC}^2)}\right]}, 
\end{equation}
with maximum at $z_{\rm GC}=3.2$, a spread of $\sigma_{\rm GC}=1.5$ \citep{mapelli2022}. The fiducial normalization constant is $\mathcal{B}_{\rm GC}=2\times{}10^{-4}\,{}{\rm M}_{\odot}\,{}{\rm Mpc}^{-3}\,{}{\rm yr}^{-1}$, consistently with \cite{el-badry2019} and \cite{reina-campos2019}.

We calculate the merger efficiency $\eta{}(Z)$ as:
\begin{equation}\label{eq:eta_dyn}
\eta{}(Z) = \frac{\mathcal{N}_{\rm merg,\,{}sim }(Z)}{\mathcal{N}_{\rm sim}(Z)}\,{}\frac{\mathcal{N}_{\rm BH}(Z)}{M_\star{}(Z)},
\end{equation}
where, for a given Z, $\mathcal{N}_{\rm merg,\,{}sim }(Z)$ is the number of simulated merging BHs, $\mathcal{N}_{\rm sim}(Z)$ is the number of simulated BHs, $\mathcal{N}_{\rm BH}$ is the total number of BHs associated with a given metallicity and $M_\star{}(Z)$ is the total initial stellar mass for a given metallicity $Z$. 
We estimate $M_\star{}(Z)$ as the sum of the initial total mass of all simulated star clusters with a given $Z$, and $\mathcal{N}_{\rm BH}(Z)$ as the number of BHs that we expect from a stellar population following a Kroupa mass function between 0.1 and 150~M$_\odot$ \citep{kroupa2001}, assuming that all stars with zero-age main sequence mass $\ge{}20$~M$_\odot$ are BH progenitors \citep{heger2003}.

We calculate
\begin{equation}
\mathcal{F}(z',z,Z) = \frac{{\rm d}\mathcal{N}(z',z,Z) / {\rm d}t}{\mathcal{N}_{\text{TOT}}(Z)}\,{}p(z', Z),
\end{equation}
where ${\rm d}\mathcal{N}(z',z,Z)/{\rm d}t$ is the total number of BBHs that form at redshift $z'$ with metallicity $Z$ and merge at redshift $z$ per unit time, and $\mathcal{N}_{\text{TOT}}(Z)$ is the total number of BBH mergers with progenitor's metallicity $Z$. 
We evaluate the dispersion around the mass-weighted metallicity by assuming a log-normal distribution for the metallicity:
\begin{equation} \label{eq:pdf}
p(z', Z) = \frac{1}{\sqrt{2 \pi\,{}\sigma_{\rm Z}^2}}\,{} \exp\left\{{-\,{} \frac{\left[\log{(Z(z')/{\rm Z}_\odot)} - {\langle{}\log{Z(z')/Z_\odot}\rangle{}}\right]^2}{2\,{}\sigma_{\rm Z}^2}}\right\},
\end{equation}
where: 
\begin{equation}
\langle{}\log{Z(z')/Z_\odot}\rangle{}=\log{\langle{}Z(z')/Z_\odot\rangle{}}-\frac{{\ln(10)}\,{}\sigma_{\rm Z}^2}{2}.
    \label{eq:average_Z}
\end{equation}
We take the average metallicity $\langle{}Z(z)/Z_\odot\rangle$ from \cite{madau2017} and assume a standard deviation $\sigma{}_Z=0.3$ \citep{mapelli2022}.  
We refer the interested reader to \cite{mapelli2022} for all the details of the calculation and the underlying assumptions.

\subsection{Description of the runs} \label{sec:runs}

For our GC models, we consider seven metallicities: $Z=0.0002$, 0.0008, 0.002, 0.004, 0.006, 0.012, and 0.02. We perform a denser sampling between 0.002 and 0.02, where the metallicity variation has a stronger impact on the 1g BH distribution. 

\begin{figure*}
\begin{center}
\includegraphics[width=\textwidth]{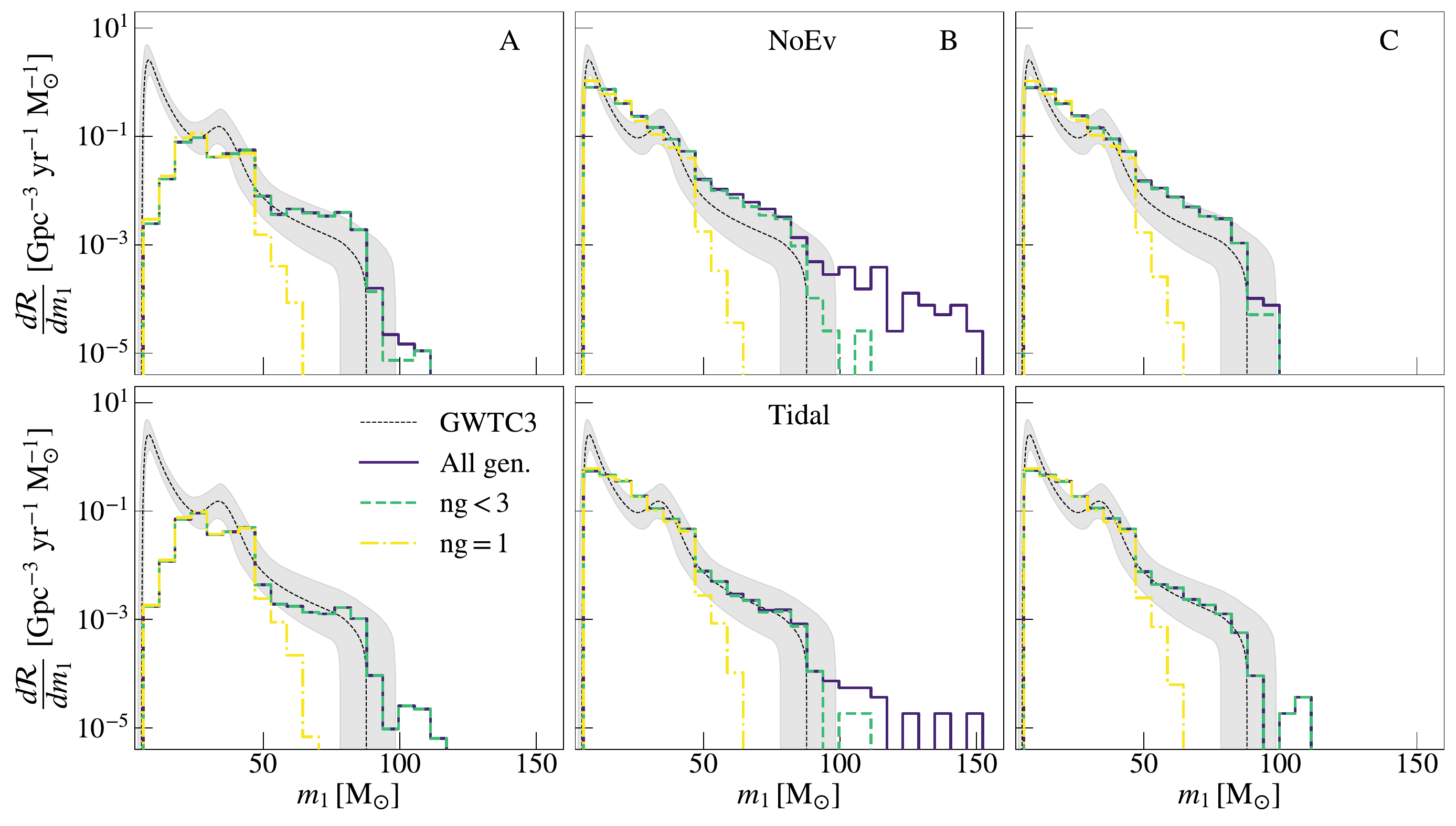}
\caption{Primary BH mass differential merger rate density in the Local Volume, for models A (left), B (center), and C (right), for the NoEv (upper panels) and Tidal (lower panels) cases. We show the mass distribution resulting from 1g mergers (yellow dash-dotted line), 1g and 2g mergers (green dashed line), and from all BBH generations together (purple solid line). The black dashed line is the median value of the \textsc{power law + peak} model, as inferred from GWTC-3 \protect\citep{abbottGWTC3population}. The grey shaded areas are the corresponding 90\% credible intervals. In this and the following Figures, the black line and shaded grey areas are shown only for a qualitative comparison, because our simulations do not assume the parametric models adopted by the LVK. 
}\label{fig:m1_inferred}
\end{center}
\end{figure*}

For each metallicity, we run three sets of $N=5 \times 10^6$ BBHs, corresponding to the mass sampling criteria introduced in Sections \ref{sec:1gBBH} and \ref{sec:Ng}. 
In the following, we refer to these three models as
\begin{itemize}
\item{\bf model A:} we draw the primary and secondary mass of 1g BBHs according to sampling A (Sect.~\ref{sec:1gBBH}),   and we sample the secondary mass of hierarchical BBHs according to the ng1g criterion (Sect.~\ref{sec:Ng});
\item{\bf model B:} we draw the primary and secondary mass of 1g BBHs according to sampling B (Sect.~\ref{sec:1gBBH}), and we sample the secondary mass of hierarchical BBHs according to the ngng criterion (Sect.~\ref{sec:Ng}), i.e. we assume that the secondary can be 
a ng BH;
\item{\bf model C:} we draw the primary and secondary mass of 1g BBHs according to sampling B (Sect.~\ref{sec:1gBBH}), and we sample the secondary mass of hierarchical BBHs according to the ng1g criterion (Sect.~\ref{sec:Ng}), i.e. we assume that the secondary is always a 1g BH.
\end{itemize}

We consider three evolutionary cases for the host GCs: \begin{itemize}
\item{\bf NoEv:} absence of star cluster evolution,
\item{\bf Evol:} star cluster evolution in isolation,  
\item{\bf Tidal:} star cluster evolution in presence of a tidal field.
\end{itemize}
To highlight the impact of  GC evolution, for each evolutionary case we set the same initial seed of our random generations. 
In total, we ran 189 different models, corresponding to  $9.45\times 10^8$ 1g BBHs. Table \ref{tab:cases} reports the details of our models.

\section{Results} \label{sec:results_fc}

\subsection{Impact of GC evolution} \label{sec:impactGCev}
Figure~\ref{fig:m1_ng} shows the impact of GC evolution on the primary BH mass distribution for different BBH generations. 
Independently of the selected mass sampling, GC evolution quenches hierarchical mergers at the third generation. This is at least one generation earlier than what predicted by non-evolving GC models.
Only in model B, GCs can still produce 4g mergers, but their efficiency is reduced by a factor of $\sim 50$ with respect to models where the GC evolution is not taken into account. The maximum primary BH mass is $\sim 150 \, \msun$ if only ng1g merger are considered, while for model B it can be as high as $\sim 300 \, \msun$.

The process that mainly quenches the hierarchical chain is the internal GC relaxation. Cluster expansion, powered by the central BH sub-system, reduces the GC central density and, in turn, the dynamical hardening rate. 
GC expansion and mass loss also lower the cluster escape velocity, favoring the ejection by relativistic recoil at merger. For 1g BBH mergers, the percentage of ejected BH remnants increases from 83\% to $92$\% when GC evolution is implemented. Only in GCs with initial $v_{\rm esc} \gtrsim 80 \, \mathrm{km \, s^{-1}}$ this percentage falls below 80\%.

As shown in Fig. \ref{fig:m1_ng}, the external tidal field does not affect the hierarchical assembly significantly.  
However, we point out that our implementation is quite conservative: it does not account for disc and bulge shocking, and for possible encounters with molecular clouds, which can significantly enhance star cluster dissolution \citep{gieles2006}. Furthermore, here we only consider GCs at 8 kpc galactocentric distance from the centre of a Milky-Way like galaxy.

\subsection{Impact of the pairing function}
The BBH mass pairing criterion deeply affects the mass distribution of BBH mergers (Fig.~\ref{fig:m1_ng}). 
In our model A, $m_1$ is  skewed towards the maximum 1g BH mass, and there are almost no mergers with $m_1 \lesssim 15 \, \msun$. In contrast, models B and C display a smoother primary mass distribution, extending down to the minimum BH mass in our models (5 M$_\odot$). 

Figure \ref{fig:mass_ratios} compares the mass ratio distributions at different generations. 
At 1g, all the mass samplings result in similar distributions, with a monotonic trend and a strong preference for $q=1$. 
For higher generations, the mass ratio distribution displays significant differences. 
In model A, the ng secondary mass is likely to be $m_2 \sim \max({m_{\rm 1,1g}})$, because this sampling favours the most massive BHs. Since $m_{\rm 1,2g} \sim 2 \max({m_{\rm 1,1g}})$, the 2g mass ratio distribution peaks at $\sim 0.5$, and the 3g distribution spreads around $1/3$. 
In model B, the mass ratios display similar distributions at all generations, since $m_2$ is determined from eq. \ref{eq:oleary}. 
In model C, the distribution shows a flatter trend after the second generation, since $m_{\rm 1,3g} \gtrsim \max{(m_{\rm 1,1g})}$ and mass ratios close to 1 are no longer possible. 

We evaluate the impact of \textbf{ng1g} mergers on BH retention by comparing the fraction of ejected BH remnants for models B and C, where the BBH masses only differ for the secondary BH mass sampling at $ng>1$. For model C, the fraction of BHs ejected by the GW relativistic recoil increases by 1\% at 2g and by 4\% at 3g, indicating that the \textbf{ng1g} assumption plays only a minor role for their retention.





\subsection{Primary BH mass merger rate density in the Local Volume}
Figure \ref{fig:m1_inferred} shows the primary mass differential merger rate density in the Local Volume. 
To obtain this quantity, we integrated the merger rate density as described in Section~\ref{sec:merger_rate}, taking into account different progenitor metallicities. 
Overall, despite the large differences for low primary BH masses ($m_1\leq{}20$ M$_\odot$), all of our models (A, B, C) reproduce both the peak at $m_1\approx{30-40}$~M$_\odot$ and the high-mass tail ($>50$ M$_\odot$) reported by the LVK collaboration \citep{abbottastrophysics}.

In all of our models, the high-mass tail of the primary BH mass distribution ($m_1>40$ M$_\odot$) is almost uniquely populated by hierarchical mergers. 
The host GC evolution quenches the contribution of hierarchical mergers to the merger rate density, because it results in a less efficient hardening rate at late times.

\begin{figure*}[ht]
\begin{center}
\includegraphics[width=\hsize]{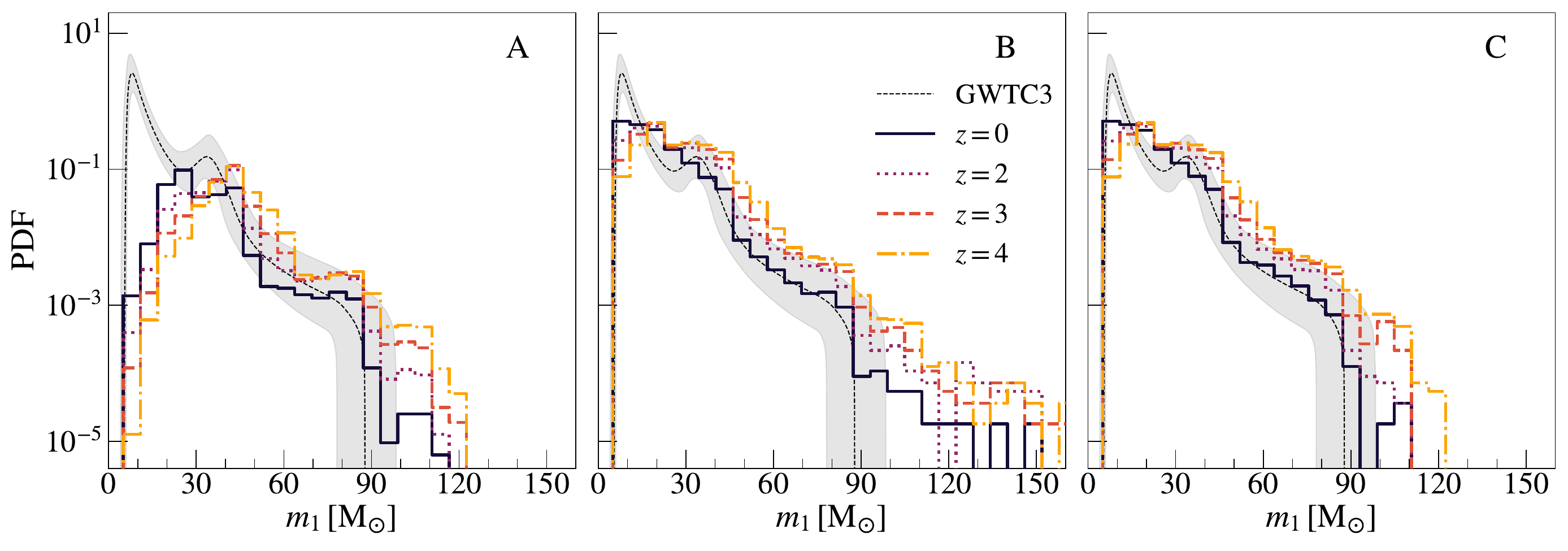}
\caption{Probability distribution function of primary BH masses at redshift $z=0$ (black solid line), $z=1$ (purple  dotted line), $z=2$ (salmon dashed line), and $z=3$ (yellow dash-dotted line) for model A (left), B (center), and C (right)  in the Tidal evolutionary case. In model B model, the distribution extends to higher masses, but the plot is truncated at $150 \, \msun$ to improve the readability of the Figure. The black dashed line is the median value of the \textsc{power law + peak} model, as inferred from GWTC-3 \protect\citep{abbottGWTC3population}. The grey shaded areas are the corresponding 90\% credible intervals. 
}\label{fig:m1_redshift}
\end{center}
\end{figure*}


The BH sampling criterion plays a key role in determining the  primary  mass distribution. 
All the models are able to account for the high-mass tail inferred by the LVK collaboration \citep{abbottGWTC3,abbottGWTC3population}, independently on the sampling criteria. 
At lower masses, instead, the considered sampling criteria differ dramatically: models B and C predict a larger BBH population at lower masses, with a main peak at $\approx 10 \, \msun$ and a monotonically decreasing trend. Also, they approximately match the second peak at $\approx 35 \, \msun$. 

In contrast, model A does not produce any peak at $\approx 10 \, \msun$. This model results in a main peak extending from $\approx 25 \, \msun$ to $\approx 50 \, \msun$, which roughly encloses the 35 M$_\odot$ peak inferred from GW observations. Here, the peak is mostly due to 1g BBH mergers. A secondary peak due to 2g BBH mergers is present at  $\approx 70-80 \, \msun$. These results are consistent with those from \cite{antonini2023}, who considered the same mass sampling criterion, but a power-law BH mass function.

This demonstrates the importance of quantifying the uncertainties related to the BH pairing function: according to models B and C, BBH mergers from GCs can account for a larger fraction of the LVK population, whereas model A hints that GCs can produce only a sub-population with primary BH mass $\gtrsim{30}$ M$_\odot$. 

\subsection{Evolution of the primary BH mass with redshift} \label{sec:redshift}
Figure~\ref{fig:m1_redshift} shows the dependence of the primary BH mass distribution with redshift. In all our models, the primary mass distribution evolves with redshift: 
BH mergers with $m_1 < 20 \, \msun$  ($m_1 >30 \, \msun$) are progressively quenched (become more common) as redshift increases. The reason for this trend is the relative contribution of different progenitor-star metallicities at different redshift (see Appendix \ref{sec:metallicity}).

In model A, the main peak of the distribution shifts from $\sim 35 \, \msun$ at $z=0$ to $\sim 50 \, \msun$ at $z=3$. Since this mass sampling is  skewed towards the maximum mass of the 1g BH spectrum, the larger contribution of sub-solar metallicities for $z>0$ produces different peaks, depending on the dominant metallicity at the considered redshift. 
In models B and C, the primary BH mass distribution for $z>0$ is no longer monotonically decreasing, but shows a depletion at $10 \, \msun$, and a main peak at $20 \, \msun$. 

Previous works  suggest the existence of a weak trend \citep{fishbach2021}, or even a strong correlation \citep{rinaldi2023} between  primary BH mass and  redshift. 
Our result indicates that  dynamically assembled BBH mergers in GCs may account for such trend. 
Finding a top-heavier BH mass function at higher redshift ($z\gtrsim{}1$) will thus be a crucial hint at the dynamical origin of these BBH mergers. Next-generation ground-based GW detectors will be able to address this point \citep{reitze2019,maggiore2020,evans2021,branchesi2023}.
\section{Discussion} 
\label{sec:discussion_fc}

\begin{figure*}[ht]
\begin{center}
\includegraphics[width=\textwidth]{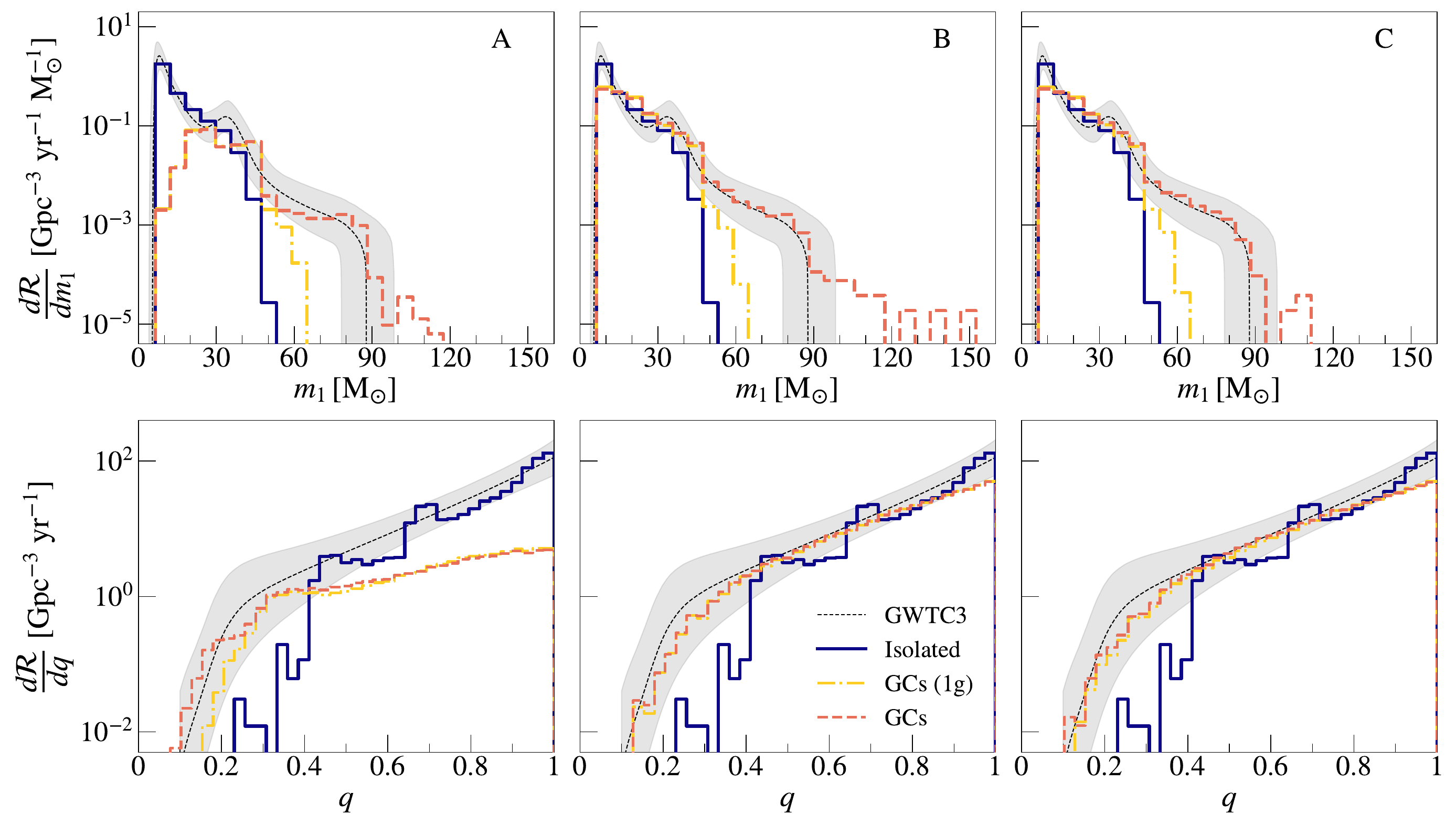}
\caption{Primary BH mass (upper panels) and mass ratio (lower panels) differential merger rate density in the Local Volume, for models A (left), B (center), and C (right) in the Tidal evolutionary case. We show the mass distribution resulting from isolated BBH mergers (blue solid line), 1g BBH mergers in GCs (yellow dash-dotted line), and all BBH mergers in GCs (including hierarchical mergers, salmon dashed line). The black dashed line is the median value of the \textsc{power law + peak} model inferred from GWTC-3 \protect\citep{abbottGWTC3population}. The grey shaded areas are the corresponding 90\% credible intervals: these are shown only for a qualitative comparison, because our simulations do not assume the parametric models adopted by the LVK. We refer to Appendix~\ref{app:Bayes} for a quantitative analysis.}  
\label{fig:mq_isoGC}
\end{center}
\end{figure*}

\begin{figure*}
\begin{center}
\includegraphics[width=\textwidth]{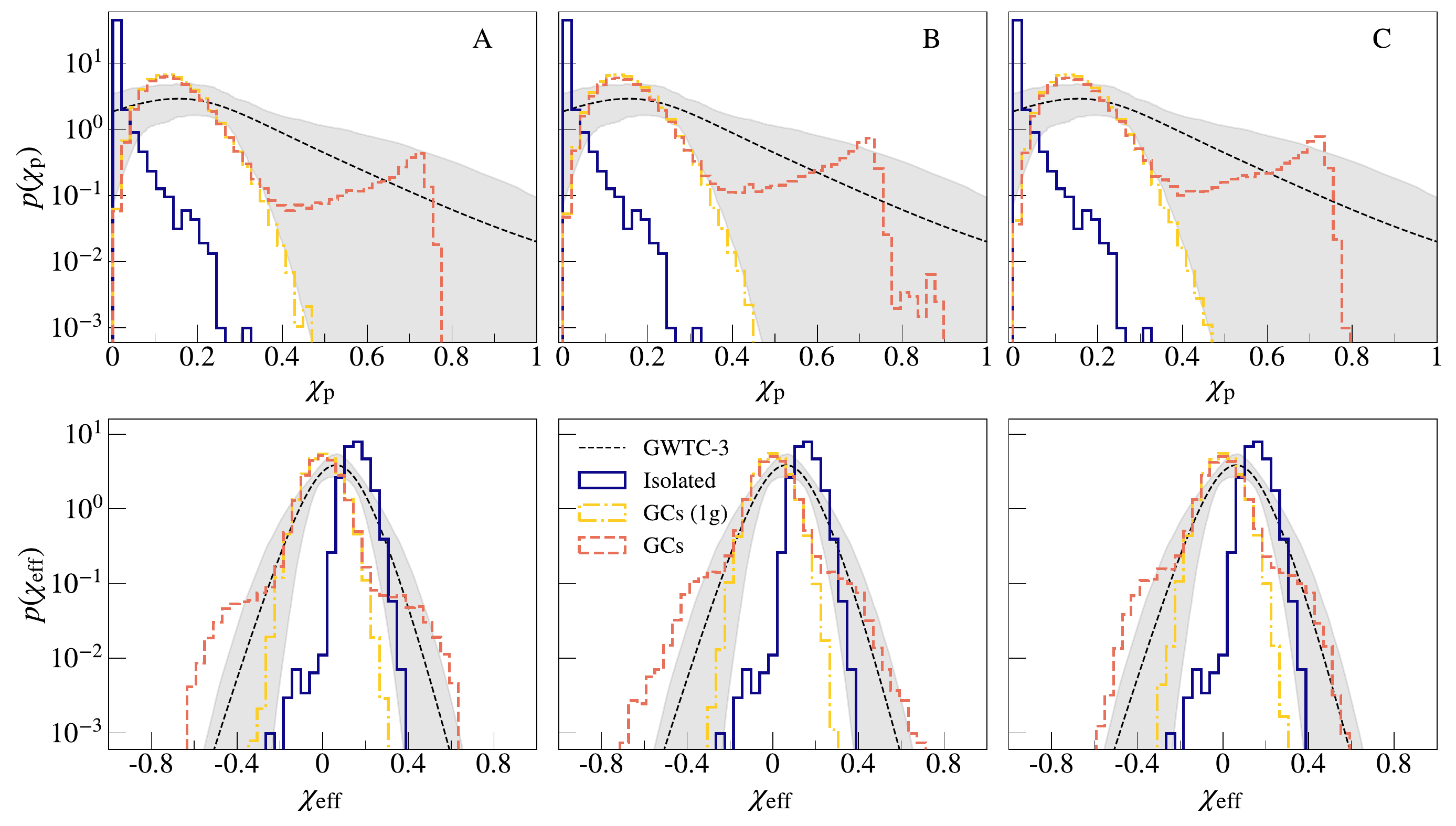}
\caption{Precessing (upper panels) and effective (lower panels) spin distribution of BBH mergers at redshift $z=0$, for models A (left), B (center), and C (right)  in the Tidal evolutionary case. We show the mass distribution resulting from isolated BBH mergers (blue solid line), 1g BBH mergers in GCs (yellow dash-dotted line), and BBH mergers in GCs, including hierarchical mergers (salmon  dashed line). The black dashed line is the median value of the Gaussian spin model inferred from GWTC-3 \citep{abbottGWTC3population}. The grey shaded areas are the corresponding 90\% credible intervals. 
}\label{fig:spins_isoGC}
\end{center}
\end{figure*}

\subsection{Comparison with isolated mergers} \label{sec:comparison_iso_GC}
Here, we compare the  primary mass, mass ratio, and spin distribution of BBHs in GCs to those predicted from isolated BBH mergers evolved with the same stellar evolution prescriptions as those used in our dynamical runs. 
We take the isolated BBH mergers from a simulation run with {\sc mobse} \citep{mapelli2017,giacobbo2018}. This simulation consists of $1.8\times{}10^8$ binary systems comprising 9 metallicities ($Z=0.0002$, 0.0008, 0.002, 0.004, 0.006, 0.008, 0.012, 0.016, 0.02). We draw the zero-age main sequence mass of the primary stars from a Kroupa \citep{kroupa2001} initial mass function between 5 and 150~M$_\odot$. We sample the initial mass ratios, orbital eccentricity and semi-major axis from the distribution by \cite{sana2012}, which are fits to observations of massive binary stars. We include core-collapse and pair-instability  supernovae as described in Sect.~\ref{sec:1gBBH} and we integrate each binary system accounting for wind mass transfer, Roche-lobe overflow, common envelope (assuming efficiency $\alpha=3$), tides, and GW  decay as described in \cite{giacobbo2018}. In order to calculate the merger rate density of isolated BBHs, we use eq.~\ref{eq:cosmorate}, assuming  cosmic star formation rate density $\psi{}(z)$  and metallicity evolution from \cite{madau2017}. We refer to \cite{santoliquido2020, santoliquido2020b} for details. 

\subsubsection{Primary mass and mass ratios}
Figure \ref{fig:mq_isoGC} shows the primary mass and mass ratio distribution  of BBH mergers from GCs and isolated binaries. As expected, primary BH masses from isolated binaries peak at $10 \, \msun$, matching the  main peak of the LVK population \citep{abbottGWTC3population}. Below $\approx{40}$~M$_\odot$, the main difference between isolated BBHs and models B and C is the relative importance of the 10 M$_\odot$ peak, which is a factor of $\sim3$ higher in isolated binaries. In contrast, model A is dramatically different from  isolated BBHs even
at low primary BH masses.

For large primary masses ($\gtrsim{40}$ M$_\odot$), all of our GC models differ from the isolated BBH model, since the latter cannot account for the high-mass tail, while the former produce high-mass BHs via hierarchical mergers.

The distribution of mass ratios (lower panel of Fig.~\ref{fig:mq_isoGC}) shows that unequal-mass BBHs are more common in our dynamical BBHs with respect to isolated BBHs. 
The mass ratios from dynamical BBHs are qualitatively consistent with the lower end of the distribution inferred from the parametric model adopted by the LVK collaboration \citep{abbottGWTC3population}.  Dynamical mergers dominate the distribution at low $q$ values.  In contrast, the distribution from the isolated channel cannot explain the LVK rate for $q<0.5$. 
In this case, the steep decrease in the mass ratio distribution at low $q$ stems from common-envelope evolution, which is responsible for BBH hardening in our models and favors the formation of equal-mass binaries. 

At high values of $q$, model A shows a flatter trend than both models B and C,  and, for $q \gtrsim 0.8$, is about one order of magnitude below the median value inferred by the LVK collaboration for the adopted parametric model  \citep{abbottGWTC3population}. 
In contrast, models B and C display an increasing trend throughout all the $q$ range and are always within the 90\% credible interval of the LVK parametric model shown in the Figure \citep{abbottGWTC3population}.

\subsubsection{Spins}
For each BBH merger, we estimate the effective spin ($\chi_{\rm eff}$) and the precessing spin ($\chi_{\rm p}$):
\begin{eqnarray}
\chi_{\rm eff}= \frac{(m_1\,{}\vec{\chi}_1+m_2\,{}\vec{\chi}_2)}{m_1+m_2}\cdot{}\frac{\vec{L}}{L},
\nonumber{}\\
\chi_{\rm p}=\frac{c}{B_1\,{}G\,{}m_1^2}\,{}\max{(B_1\,{}S_{1\perp{}},\,{}B_2\,{}S_{2\,{}\perp})},
\end{eqnarray}
where $\vec{L}$ is the orbital angular momentum of the BBH, $B_1\equiv{}2+3\,{}q/2$ and $B_2\equiv{}2+3/(2\,{}q)$. $S_{1\perp{}}$ and $S_{2\perp{}}$ are the components of the spin vectors ($\vec{S}_1$ and $\vec{S}_2$) perpendicular to the orbital angular momentum.  $\chi_{\rm eff}$ characterizes a mass-averaged spin angular momentum in the direction parallel to the binaries orbital angular momentum, while $\chi_{\rm p}$ corresponds approximately to the degree of in-plane spin, and traces the rate of relativistic precession of the orbital plane.

Figure \ref{fig:spins_isoGC} shows the distributions of $\chi_{\rm eff}$ and $\chi_{\rm p}$ from our models in the Tidal evolutionary case, compared to the inferred distributions through the Gaussian spin model presented by \cite{abbottGWTC3population}.  
The isolated channel mainly leads to the formation of low $\chi_{\rm p}$ and positive $\chi_{\rm eff}$, because of angular  momentum alignment driven by mass transfer and tidal effects. First-generation BBHs in GCs show a distribution of $\chi_{\rm eff}$ symmetric around zero, due to the assumption of isotropic spins, and a peak at $\chi_{\rm p}\approx{0.15}$. The spread of the distributions of $\chi_{\rm eff}$ and  $\chi_{\rm p}$ depend on the initial value of the parameter $\sigma_{\chi}$ of the initial Maxwellian distribution, which, in our models is set to 0.1 (Section~\ref{sec:1gBBH}).

BHs from previous mergers typically display larger spins, resulting from the sum of their parent BBH orbital angular momentum and BH spins.
This produces a spin distribution peaked at $\chi \sim 0.7$ \citep{gerosa2017,fishbach2017}, and, in turn,  a peak at $\chi_{\rm p}=0.7$. In model B, where the contribution of 3g mergers is not negligible, a tail at $\chi_{\rm p} \approx 0.8$ is present, due to the further spin increase at successive generations. Also, the high spins produced by hierarchical mergers can extend the distribution of $\chi_{\rm eff}$ to much lower and higher values. 

The population inferred from GWs is mostly consistent with slowly rotating BHs \citep{callister2022}, but the decrease at high spins is milder than what expected from 1g mergers only, according to our models. 
At the same time, high absolute values of $\chi_{\rm eff}$ from hierarchical mergers are not ruled out by current GW detections. Given the current uncertainties on spin parameters, 
additional GW detections will be crucial to 
constrain a possible secondary peak in the distribution of $\chi_{\rm p}$, due to hierarchical mergers.

\subsection{Mixing fractions}
\label{sec:mixing_fractions}

As a simple but instructive exercise, we calculate the mixing fraction between two channels: BBHs born in GCs and BBHs formed in isolated binary systems. For the latter, we use the models described in Section~\ref{sec:comparison_iso_GC}.   We neglect here the other dynamical channels for simplicity, because a  two-channel analysis allows for a more straightforward interpretation of the impact of the main model features. We refer to \cite{mapelli2022} for a more sophisticated but somewhat less transparent four-channel analysis. 

We calculate the probability of recovering the parameters $\theta=\{\mathcal{M}_{\rm chirp},\,{}q,\,{}z,\,{}\chi_{\rm eff},\,{}\chi_{\rm p}\}$ of the observed LVK events\footnote{In our formalism, $\mathcal{M}_{\rm chirp}=(m_1\,{}m_2)^{3/5}\,{}(m_1+m_2)^{-1/5}$ is the BBH chirp mass.} \citep{abbottGWTC3population} given the mixing fraction $f_{\rm GC}$ and the other hyperparameters of our models $\lambda{}$:
\begin{equation}\label{eq:mixfraction}
p(\theta{}|f_{\rm GC},\,{}\lambda{})=f_{\rm GC}\,{}p(\theta{}|{\rm GC},\lambda{})+(1-f_{\rm GC})\,{}p(\theta{}|{\rm iso},\lambda{}).
\end{equation}
According to this formalism, a mixing fraction $f_{\rm GC}=0$ ($f_{\rm GC}=1$) means that all BBHs are born from isolated binary systems (GCs). In our analysis, we consider 58 BBH events from \cite{abbottGWTC3} with probability of astrophysical origin $p_{\rm astro}>0.9$ and false-alarm rate FAR$<0.25$~yr$^{-1}$. 
We describe our hierarchical Bayesian framework in Appendix~\ref{app:Bayes}. With this approach, we also take into account  observational biases, associating  a detection probability to our simulated catalogues. This allows us to do a fair comparison between our models and LVK data. 

\begin{figure}
\begin{center}
\includegraphics[width=\hsize]{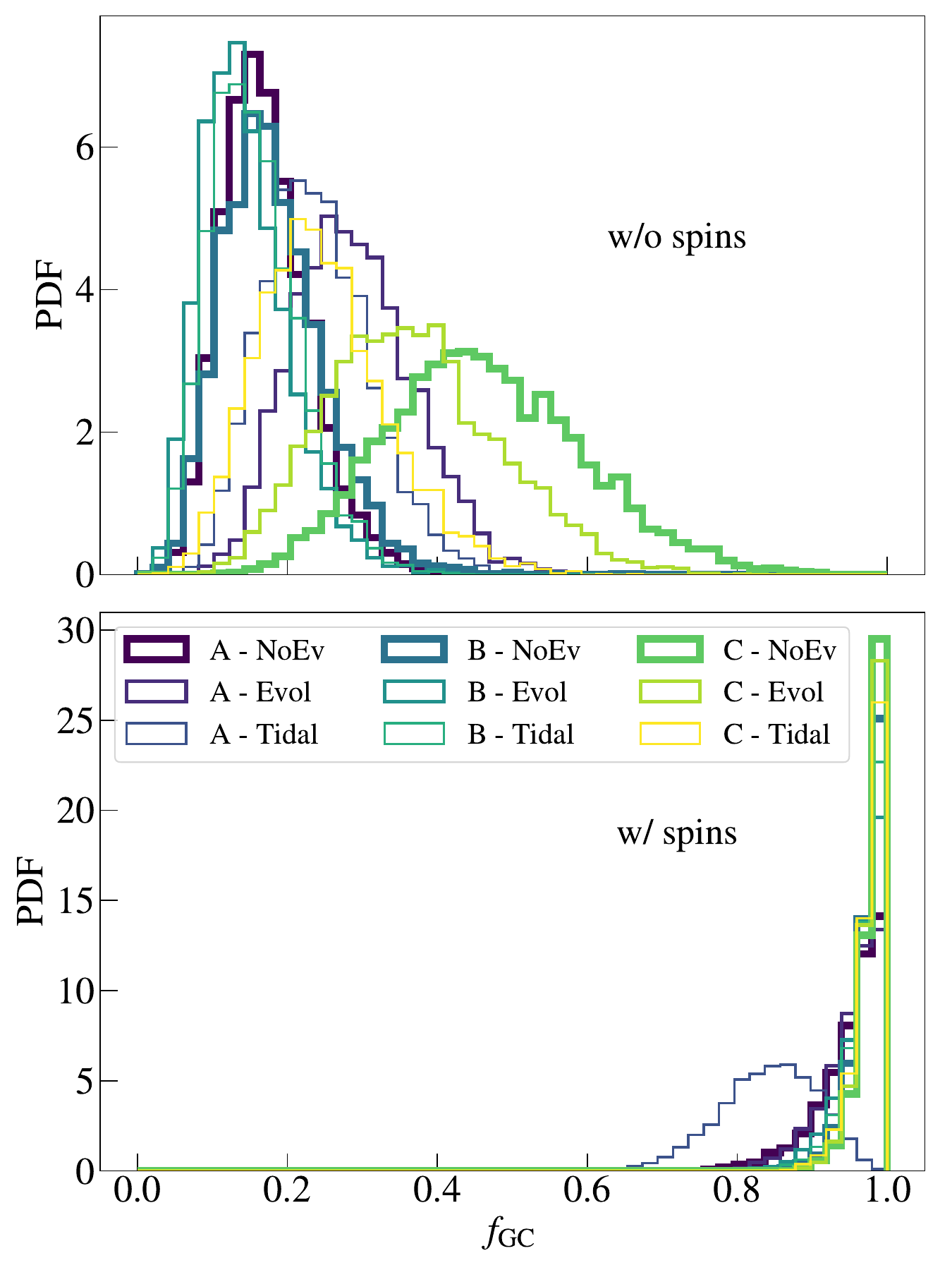}
\caption{Mixing fraction ($f_{\rm GC}$) probability distribution function. A mixing fraction $f_{\rm GC}=0$ ($f_{\rm GC}=1$) means that all BBHs are born from isolated binary systems (GCs). Upper (Lower) panel: we (do not) marginalize over the  spin parameters $\chi_{\rm eff}$ and $\chi_{\rm p}$ in our analysis. All the models presented in this paper are shown with different colours, as described in the legend.
}\label{fig:mix_frac}
\end{center}
\end{figure}

Figure~\ref{fig:mix_frac} shows the probability distribution function (PDF) of the GC mixing fraction $f_{\rm GC}$ we obtain if we marginalize (upper panel) or we do not marginalize over $\chi_{\rm eff}$ and $\chi_{\rm p}$ (lower panel). 
This Figure shows a clear predominance of the GC channel if we include the spin parameters in the calculation of the mixing fraction, almost regardless of the GC model we assume. The main reason is that our isolated binary model tends to produce spins that are too much aligned with the orbital angular momentum in order to account for the majority of the LVK data (see Fig.~\ref{fig:spins_isoGC} and Appendix~\ref{app:Bayes} for details). This result confirms that the evolution of spins' orientation during binary evolution is a key process (e.g., \citealt{stegmann2021,callister2021spin}). 
As shown by \cite{mapelli2022}, this result is degenerate with the initial assumption on spin magnitudes. In particular, the choice of lower 1g spin magnitudes (e.g., $\sigma_{\rm \chi}=0.01$) can lead to a higher contribution of the isolated BBHs to the mixing fraction.

If we do not include $\chi_{\rm eff}$ and $\chi_{\rm p}$ in our analysis, that is if we marginalize over them, the GC mixing fraction $f_{\rm GC}$ decreases significantly, implying an important contribution of the isolated BBH channel. The contribution of the isolated BBHs when we neglect the spins   mostly stems from the impact of  $m_1$. In fact, our isolated binary models have a preference for $m_1\sim{8-10}$ M$_\odot$, which corresponds to the main peak in the LVK population (Fig.~\ref{fig:mq_isoGC}). In contrast, such peak is suppressed in all of our dynamical models, especially model \textbf{A}. 

Model C -- NoEv  results in the largest GC mixing fraction among dynamical models if we neglect the spins (median value of $f_{\rm GC}\approx{0.4}$ for model C -- NoEv), whereas models A and B cluster at lower values of $f_{\rm GC}$.  This difference is more prone to stochastic fluctuations in our simulations. However, we note that model C -- NoEv better reproduces the events with both a large value of chirp mass $\mathcal{M}_{\rm chirp}$ and a low value of $q$ in the LVK data (Appendix~\ref{app:Bayes}). Overall, the differences between models Evol/NoEv/Tidal are negligible in the analysis of the mixing fractions.

\section{Summary} \label{sec:conclusions_fc}

We have investigated the process of hierarchical mergers in globular clusters (GCs), accounting for the main physical processes that drive  GC evolution, namely mass loss by stellar evolution, expansion powered by a central BH sub-system, and tidal stripping.  We also bracket the uncertainties related to BH pairing functions 
on the predicted primary BH mass, mass ratio, and spin distribution.

We summarize our main results as follows.

\begin{itemize}
\item GC  expansion quenches the hierarchical BH assembly already at the third generation. 
Our models predict a maximum BH mass of $\sim140 \, \msun$ if the mergers only involve first-generation secondary BHs, 
and $\sim300 \, \msun$ if we include the possibility of ng secondary BHs. 

\item In all of our models, the hierarchical mergers dominate the primary BH mass distribution for $m_1 > 50 \, \msun$, and can reproduce the high-mass tail inferred from GW events. 

\item  The adopted mass pairing function deeply affects the low-mass end of the primary BH mass distribution. 
Model A suppresses the formation of low-mass BBHs, resulting in a main peak around $m_1=35 \, \msun$.  
In contrast, models B and C produce a peak in the primary BH mass function at $m_1\sim{10}$ M$_\odot$ and then a monotonic decrease for larger masses. 

\item The primary BH mass distribution  evolves with redshift in all of our GC models, with a larger contribution from mergers with $m_1 \geq 30 \, \msun$ at $z\geq{}2$, while low-mass BBH mergers are quenched at high redshift. This effect stems from the interplay between GC dynamics and stellar metallicity evolution.

\item Overall, despite the large differences for low primary BH masses ($m_1\leq{}20$ M$_\odot$), all of our models (A, B, and C) match both the peak at $m_1\approx{30-40}$~M$_\odot$ and the high-mass tail ($>50$ M$_\odot$) reported by the LVK collaboration \citep{abbottGWTC3population}. This confirms the importance of star-cluster dynamics and hierarchical mergers to produce the most massive BBH mergers observed with LVK.

\item We compared the primary mass, mass ratio, and spin distribution of dynamical BBHs in GCs to those derived from isolated BBH mergers. Isolated binaries result in a contribution $\sim 3$ times higher than  
models B and C at $10 \, \msun$,   
matching the main peak from LVK data. In contrast, dynamical mergers in GCs are necessary to reproduce the events with $m_1 > 50 \, \msun$ and $q<0.5$. Also, hierarchical mergers imprint a secondary peak in the $\chi_{\mathrm{p}}$ distribution and contribute to the positive and negative wings of the $\chi_{\mathrm{eff}}$ distribution. 

\item We calculated the mixing fraction between dynamical BBHs in GCs and isolated BBHs. Our predictions are very sensitive to the spin parameters, which favour a large fraction of dynamical BBH mergers in GCs ($f_{\mathrm{GC}}>0.6$). 

\end{itemize}

Our analysis confirms the importance of hierarchical mergers to match the high-value tails of the primary BH mass ($m_1\geq{}50$ M$_\odot$) and  precession spin distributions ($\chi_{\rm p}\sim{0.7}$). We predict that the mass of BBH mergers in GCs increases with redshift, regardless of the uncertainties 
related to the BBH pairing function. 
The ongoing fourth observing run of the LVK and next-generation GW detectors will be an important test-bed for our  key predictions.

\section*{Data Availability}
\fastcluster{} is an open-source code available at \href{https://gitlab.com/micmap/fastcluster_open}{this link}. The latest public version of {\sc mobse } can be downloaded from \href{https://gitlab.com/micmap/mobse_open}{this repository}. {\sc clusterBH} is available at \href{https://github.com/mgieles/clusterbh}{this link}. 
The data underlying this article will be shared on reasonable request to the corresponding authors.

\section*{Acknowledgements}
We thank the anonymous referee for their comments, which helped us to improve this manuscript. 
We thank Mark Gieles and Fabio Antonini for sharing {\sc clusterBH} and for  productive discussions. MM, ST, CP, MD and MPV acknowledge financial support from the European Research Council for the ERC Consolidator grant DEMOBLACK, under contract no. 770017. ST, MM and MPV also acknowledge financial support from the German Excellence Strategy via the Heidelberg Cluster of Excellence (EXC 2181 - 390900948) STRUCTURES. MD also acknowledges financial support from Cariparo foundation under grant 55440. MAS acknowledges funding from the European Union’s Horizon 2020 research and innovation programme under the Marie Skłodowska-Curie grant agreement No.~101025436 (project GRACE-BH). MCA acknowledges financial support from the Seal of Excellence @UNIPD 2020 program under the ACROGAL project.



\bibliographystyle{aa} 
\bibliography{references.bib} 




\begin{appendix}
\section{Star cluster evolution} \label{sec:coupling}

In the following, we summarize the description of GC evolution we included in our treatment. We use the model presented by \cite{antonini2020} and the code \clusterbh{}.


\subsection{Stellar mass loss}
Stellar mass loss by stellar winds and supernova explosions, and the subsequent GC expansion are modeled as 
\begin{equation}
\dot{M}_{\star,\mathrm{sev}} = 
\begin{cases}
0  & t < t_{\mathrm{sev}}, \\
- \nu \, \dfrac{M_{\star}}{t}  & t \geq t_{\mathrm{sev}},
\end{cases}
\end{equation}
\begin{equation}
\dot{r}_{\mathrm{h,sev}} = - \frac{\dot{M}_{\star,\mathrm{sev}}}{M_{\mathrm{tot}}} r_{\mathrm{h}},
\end{equation}
where $M_\ast$ is the total stellar mass, $t_{\mathrm{sev}}$ is the time-scale for stellar evolution, $r_{\rm h}$ is the half-mass radius, and $\nu = 0.07$ \citep{antonini2020}. 
We calibrate $t_{\mathrm{sev}}$ to match the trend from direct $N-$body simulations, as described in Sect.~\ref{sec:test}.

\begin{figure*}
\begin{center}
\includegraphics[width=\hsize]{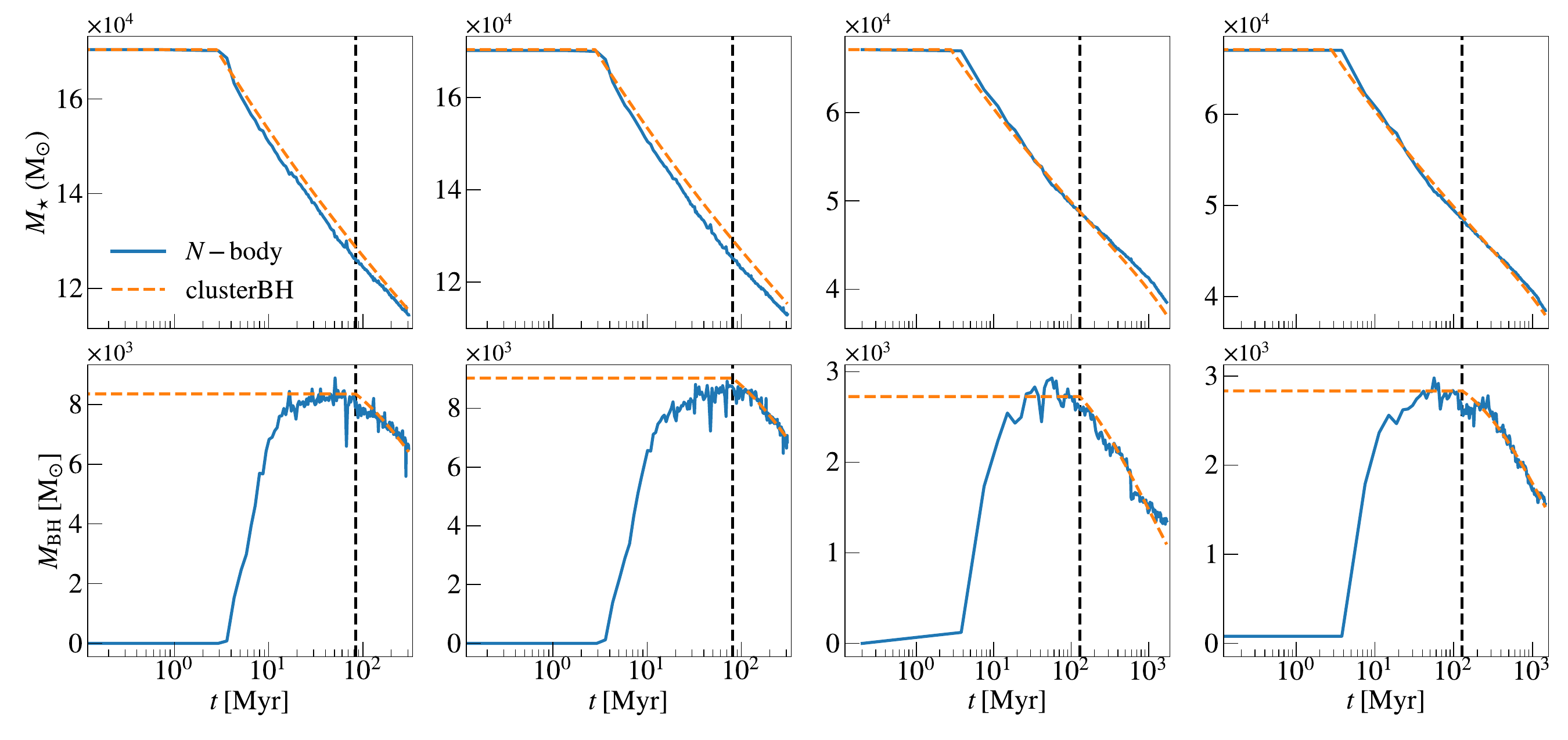}
\caption{Evolution of $M_{\star}$ (top) and $M_{\mathrm{BH}}$ (bottom), in the $N-$body simulations (blue, solid line) presented in \protect\cite{arcasedda2023d1} and in this work (orange dashed line). Different columns show the evolution of 
some representative $N-$body simulations. The two left-hand panels show the evolution of a GC with $N=3\times 10^5$ stars, while
the two right-hand panels show the evolution of GCs with $N=1.2\times 10^5$ stars. 
The black vertical dashed line is the time of the GC core collapse. }\label{fig:comparison}
\end{center}
\end{figure*}

\subsection{Two-body relaxation} \label{sec:relaxation_fc}
We consider the onset of balanced evolution as the moment of the GC core collapse \citep{antonini2020}:
\begin{equation} \label{eq:tcc}
    t_{\rm{cc}} = 3.2 \, t_{\rm{rh, \psi}}, 
\end{equation}
where 
$t_{\rm{rh,\psi}}$ is the half-mass relaxation time-scale \citep{spitzer1971}:
\begin{equation}
    t_{\rm{rh, \psi}} = 0.138 \sqrt{\frac{M_{\rm{tot}}\,{} r^3_{\rm{h}}}{G}} \frac{1}{\langle m \rangle \,{}\psi \,{}\rm{\ln{\Lambda}}}.
\end{equation}
Here, $G$ is the gravitational constant, $M_{\rm tot}$ is the total cluster mass (both stars and black holes), $\langle m \rangle$ is the mean mass within the cluster (including both stars and black holes), and $\rm{\ln{\Lambda}}=10$ is the Coulomb logarithm. The parameter $\psi$ is the heat conduction efficiency due to the mass function. 

Following \cite{spitzer1971}, for a two-component model (the two components being black holes and stars):
\begin{equation}
\psi = \left(m^{3/2}_{\star}\,{}M_{\star}+m_{\rm{BH}}^{3/2}\,{}M_{\rm{BH}}\right) \, N^{3/2}_{\rm{tot}} / M^{5/2}_{\rm{tot}},
\end{equation}
where $N_{\rm{tot}}$ is the total number of stars within the cluster. In our runs the mean stellar mass $m_{\star}=0.6 \, \msun$, following a \cite{kroupa2001} initial mass function, while $m_{\rm{BH}}$ is the mean mass of first-generation BHs (see Sect. \ref{sec:fastcluster}), and thus depends on the metallicity. We set the total BH mass, $M_{\rm{BH}}$, based on the BH mass fraction from direct $N-$body models, as described in Sect. \ref{sec:test}.

When the GC reaches a state of balanced evolution, 
the expansion rate is given by
\citep{antonini2020}:
\begin{equation}
    \dot{r}_{\mathrm{h,rlx}} = \zeta \,{}\frac{r_{\mathrm{h}}}{t_{\mathrm{rh,\psi}}} + 2 \,{}\frac{\dot{M}_{\mathrm{tot}}}{M_{\mathrm{tot}}} \,{}r_{\rm{h}}.
\end{equation}
Here, $\dot{M}_{\rm{tot}}=\dot{M}_{\star}+\dot{M}_{\rm{BH}}$, and $\dot{M}_{\rm{BH}}$ is the BH mass-loss rate due to dynamical ejections within the core \citep{breen2013a}:  
\begin{equation}
\dot{M}_{\mathrm{BH}} = \begin{cases}
0, & t < t_{\mathrm{cc}} \; \mathrm{or} \; M_{\mathrm{BH}}=0  \\
 -\beta \dfrac{M_{\mathrm{tot}}}{t_{\mathrm{rh, \psi}}} & t \geq t_{\mathrm{cc}} \; \mathrm{and} \; M_{\mathrm{BH}}>0,
\end{cases}
\end{equation}
where $\beta = 3 \times 10^{-3}$ (\citealp{antonini2020}).

\subsection{Tidal mass loss} \label{sec:tidal}
Tidal stripping from the host galaxy brings about additional stellar-mass loss to the cluster. Following \cite{gieles2011a}, we model the tidal mass-loss rate from a point-mass galaxy under the assumption of circular orbits as:
\begin{equation}
 \dot{M}_{\star, \rm{tid}} = - \xi_{\rm{J}} \,{}\frac{ M_{\rm{tot}}}{ t_{\rm{rh,\psi=1}} }.
\end{equation}
Here, we set the tidal mass loss timescale to the relaxation time for $\psi=1$ \citep{gieles2011a}, because this process takes place at the tidal boundary, where almost no BHs are present. The parameter $\xi_{\rm{J}}$ embodies the impact of different tidally-filling regimes on the cluster mass loss:
\begin{equation}
\xi_{\rm{J}} = \frac{3}{5} \zeta \left(\frac{r_{\rm{h}}/r_{\rm{J}}}{\left[r_{\rm{h}}/r_{\rm{J}}\right]_1}\right)^{3/2}.
\end{equation}
Here, $[r_{\rm{h}}/r_{\rm{J}}]_1$ is defined as in \cite{gieles2011a} and $r_{\rm{J}}$ is the Jacobi (tidal) radius \citep{gieles2008}:
\begin{equation}
    r_{\rm{J}} = \left( \frac{G}{3 \omega^2} \right)^{1/3} M_{\rm{tot}}^{1/3},
\end{equation}
where $\omega= V_{\rm{C}} \, R_{\rm{G}}^{-1} $ depends on the cluster galactocentric distance $R_{\rm{G}}$ and its circular velocity $V_{\rm{C}}$ \citep{gieles2008}. For our runs, we consider $R_{\mathrm{G}}=8$ kpc and $V_{\mathrm{C}}= 220 \, \mathrm{km \, s^{-1}}$, which correspond to a solar neighbourhood-like static tidal field.


The combined stellar mass loss, including stellar evolution and tidal stripping, is:
\begin{equation}
\dot{M}_{\rm{\star}} = 
\begin{cases}
\dot{M}_{\rm{\star}, tid}  & t < t_{\mathrm{sev}}, \\
\dot{M}_{\rm{\star,tid}} + \dot{M}_{\rm{\star,sev}} & t \geq t_{\mathrm{sev}}. 
\end{cases} 
\end{equation}

We also introduced the possibility to model the galactic rotational curve at different galactocentric distances, based on a \cite{dehnen1993} profile:
\begin{equation}
    V_{C} = G M_{\rm{g}} \frac{R_{\rm G}^{2- \gamma}}{(R_{\rm G}+r_{\rm{s}})^{3-\gamma}},
\end{equation}
Where $M_{\rm{g}}$ is the galaxy total mass and $r_{\rm{s}}$ is the galaxy length scale. Although this feature was not considered for this work, it will be a useful tool for investigating the evolution of GC populations in galaxies.



\subsection{Calibration of \texorpdfstring{$t_{\rm{sev}}$}{tsev} and \texorpdfstring{$M_{\rm{BH}}$}{MBH}}\label{sec:test}
The stellar evolution timescale, $t_{\rm{sev}}$, and the BH mass within the cluster, $M_{\rm{BH}}$, depend on the the details of stellar evolution and BH ejection by  supernova kicks, respectively. Here, we tuned these parameters to reproduce the evolution of the stellar and BH mass 
of the direct $N-$body simulations by \cite{arcasedda2023d1,arcasedda2023d3,arcasedda2023d2}, following a similar approach to \cite{antonini2020}.  

As shown in Figure \ref{fig:comparison}, our semi-analytic implementation can trace the trend of stellar mass loss. By setting $t_{\rm{sev}}=3$ Myr, we capture the early change in the stellar mass slope due to the onset of supernova explosions. For this reason, in the following we will consider $t_{\rm{sev}}=3$~Myr. 

In the $N-$body models, the total BH mass within the half-mass radius initially increases because the most massive stars progressively give birth to BHs, 
and peaks at $t_{\mathrm{cc}}$. Since then, the total BH mass decreases because of dynamical ejections within the BH sub-system. For this comparison, we set the initial $M_{\mathrm{BH}}$ to the value given by the $N-$body simulations at $t_{\mathrm{cc}}$,  which ranges from 0.042 to 0.05. 
In our runs, we will assume an initial BH mass fraction of $0.045$.

\subsection{Coupling between \clusterbh{} and \fastcluster{}}

In this Section, we describe how we interfaced the two semi-analytic codes \clusterbh{} \citep{antonini2020} and \fastcluster{} \citep{mapelli2021}. First, we generate the GC masses and densities and calculate the time of core collapse (eq.~\ref{eq:tcc}). We draw the 1g BBH masses, spins and orbital properties as described in Sect. \ref{sec:1gBBH}. For each metallicity, we set in both \fastcluster{} and \clusterbh{} the value of $m_{\rm BH}$ to the mean BH mass in the {\sc mobse} catalog.

We evolve the host GC until the time of core collapse, if BHs are not ejected by their natal kick. To ensure efficiency along with accuracy, we implemented  a Runge-Kutta fourth order integrator and an adaptive time-step in \clusterbh{}. 
Specifically, we increase (reduce) the time-step by a factor of 2 (10) if the percentage change of the integrated quantities ($M_{\star}$, $M_{\rm BH}$ or $r_{\rm hm}$) between two time-steps falls below $0.1\%$ (exceeds 1\%). 
After BBH formation, we integrate the GC mass, radius, density and velocity dispersion, along with the BBH eccentricity and semi-major axis (see eq.~16 in \citealp{mapelli2021}).
Since the evolution timescales of the BBH and the host GC generally differ by several orders of magnitude, we have implemented two separate adaptive time-step schemes for integrating the BBH hardening and the GC evolution, respectively.

We stop GC evolution if the BH is ejected from the cluster (e.g. as a consequence of three-body interactions or relativistic recoil) or when the integration reaches the Hubble time. If the GC evaporates, or no BHs are retained within the cluster, we continue the BBH evolution as an isolated binary, as described in \cite{mapelli2021}. 
Our implementation delivers an efficient integration of both the GC and the BBH properties, and enables us to evolve $\sim 3 \times 10^5$ BBHs in GCs per day with a single core.

\section{Impact of the initial GC mass and density on our results} \label{app:M0_rho0}
The initial GC mass and density distributions we have assumed in the main text  (Sect. \ref{sec:clusters}) 
match the observed properties of current Milky Way GCs \citep{harris2013,vandenberg2013}. However, GCs might have been more massive and denser at their formation time \citep[e.g.,][]{vesperini1998}.  In the following, we estimate the impact of  GC initial conditions on the BBH merger rate density, by considering larger initial GC masses and densities. In particular, we assume that the initial median value of the GC mass (density) distribution is two (ten) times as large as the values considered in Sect. \ref{sec:clusters}. 

\begin{figure*}
\begin{center}
\includegraphics[width=\hsize]{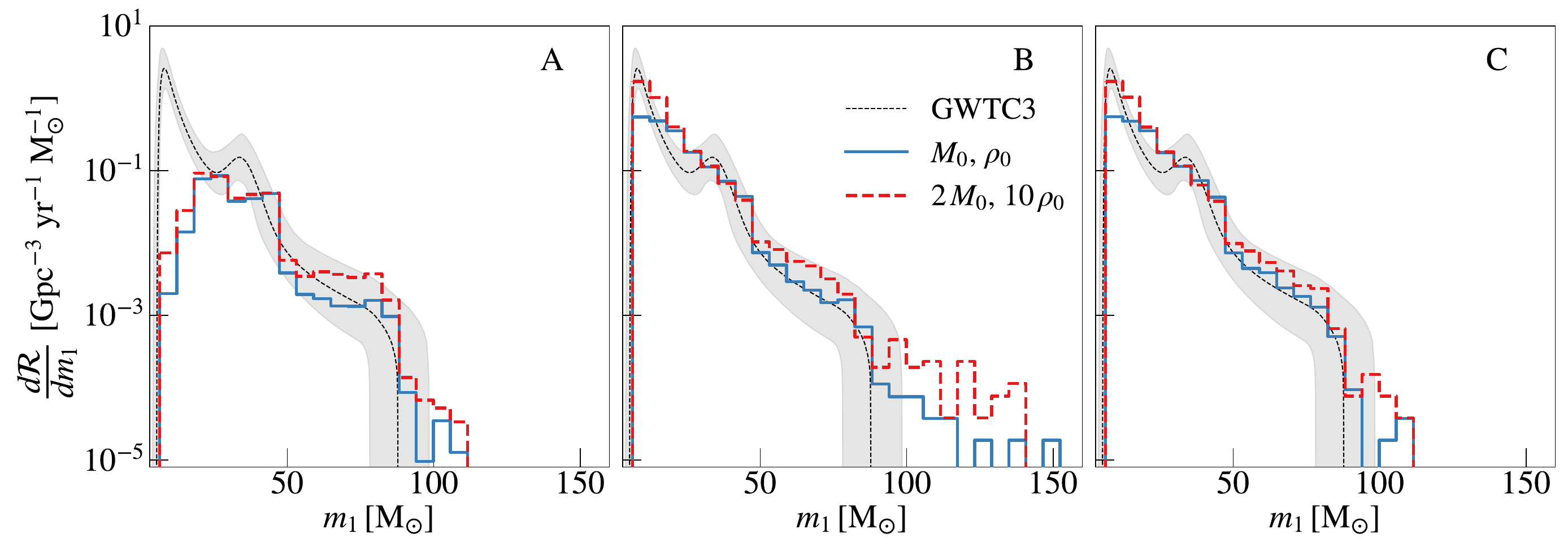}
\caption{
 Primary BH mass merger rate density for models A (left), B (center), and C (right) in the Tidal evolutionary case. The GC masses and densities are initialized as described in Sect. \ref{sec:clusters}, ($M_0, \rho_0$, blue solid line), and with mass (density) distributions that are shifted to larger values by a factor 2 (10, red dashed line). The black dashed line is the median value of the \textsc{power law + peak} model inferred from GWTC-3 \protect\citep{abbottGWTC3population}. The grey shaded areas are the corresponding 90\% credible intervals. 
}\label{fig:m1_mass_rho}
\end{center}
\end{figure*}

Figure \ref{fig:m1_mass_rho} shows that larger GC masses and densities have a mild impact on the primary BH mass.  
As expected, denser and more massive clusters produce a higher BBH merger rate density, because of a higher BH retention fraction. 
In models B and C, the merger rate of low-mass BHs is  enhanced, because supernova kicks are less efficient in ejecting them from more massive GCs. In contrast, the mass sampling presented in model A is only marginally affected by this effect, because only massive BHs tend to pair-up efficiently with this mass sampling.

The longer cluster lifetime and the shorter evolution timescales also enhance the production of hierarchical mergers. However, also in this case we find a minor contribution from $\mathrm{ng} \geq 3$ mergers at redshift $z=0$, especially for the $\mathrm{ng1g}$ cases (models A and C).

\section{Hierarchical BH mass function as a function of metallicity} \label{sec:metallicity}

\begin{figure*}
\begin{center}
\includegraphics[width=\hsize]{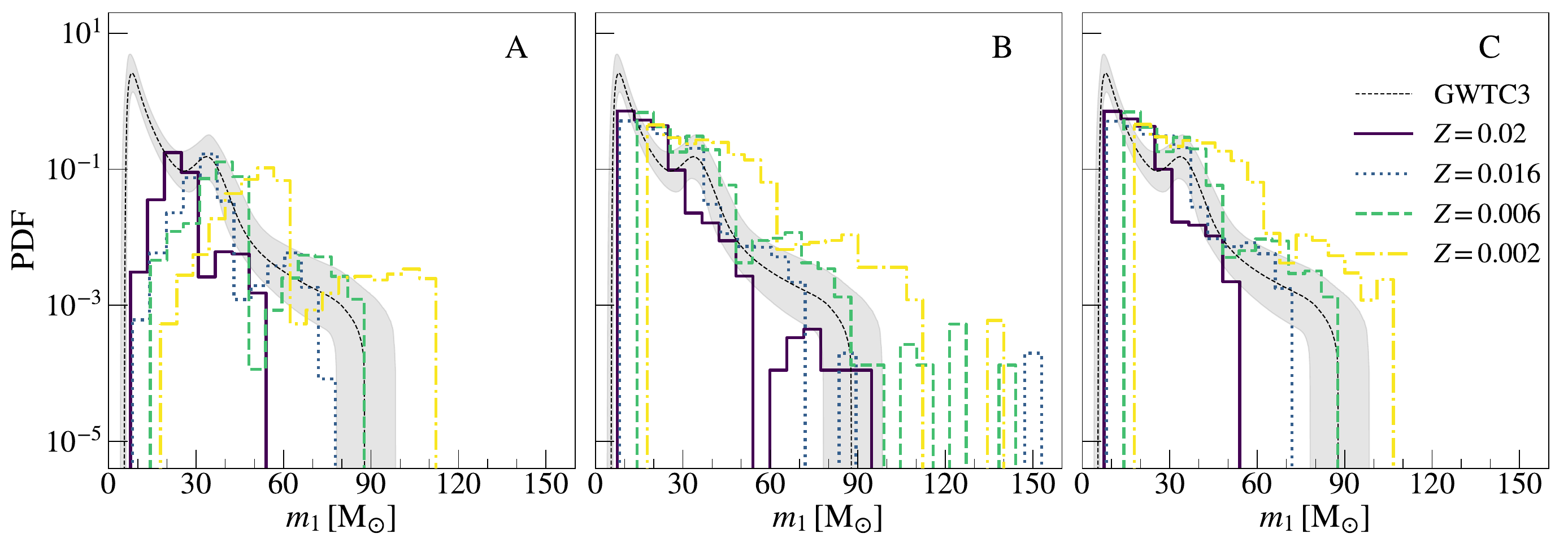}
\caption{Probability distribution function of primary BH masses in GCs, for models A (left), B (center), and C (right) in the Tidal evolutionary case. We consider different progenitor metallicities: $Z=0.02$ (purple solid line), $Z=0.016$ (blue dotted line), $Z=0.006$ (green dashed line), and $Z=0.002$ (yellow dash-dotted line). The black dashed line is the median value of the \textsc{power law + peak} model inferred from GWTC-3 \protect\citep{abbottGWTC3population}. The grey shaded areas are the corresponding 90\% credible intervals. 
}\label{fig:m1_metallicity}
\end{center}
\end{figure*}

Figure \ref{fig:m1_metallicity} shows the primary BH mass distribution produced by BBH mergers at different metallicities.  
In model A, the main peak  shifts with the maximum mass of  first-generation BHs. Also, for each metallicity, this pairing function predicts a secondary peak due to hierarchical mergers at $\sim 2 \max({m_{\rm 1,1g}})$.

In models B and C, the main peak at $10 \, \msun$ is mainly due to BBH mergers at solar ($Z=0.02$) and slightly sub-solar metallicity ($Z=0.016$), while lower metallicities extend the mass distribution toward larger masses. 

As discussed in Sect. \ref{sec:redshift}, the relative contribution of different metallicities results in a variation of the BH primary mass distribution with redshift. 
 In models~B and C, primary BH masses $m_1<20 \, \msun$ become less and less common at $z\geq2$, because metal-rich progenitors ($Z=0.02,0.016$) also become less frequent at high redshift.  
 
 In model~A, the main peak of the primary BH mass shifts from $\sim 35 \, \msun$ to $\sim 50 \, \msun$ as a function of redshift, because metal-poor stars become more and more common at high redshift.

\section{Bayesian hierarchical analysis} \label{app:Bayes}

To compare our models against GW events in the third-GW transient catalogue, we use a hierarchical Bayesian approach, as we already described in previous work \citep{bouffanais2019,bouffanais2021,mapelli2022}. Given a number $N_{\rm obs}$ of GW observations, $\mathcal{H}=\lbrace h^{k} \rbrace_{k=1}^{N_{\rm obs}}$, described by an ensemble of parameters $\theta$, the posterior distribution of the hyper-parameters $\lambda{}$ associated with the models is described as an in-homogeneous Poisson distribution \citep{loredo2004,mandel2018}

\begin{eqnarray}\label{eq:post_hier_model}
p(\lambda{}, N_\lambda | \mathcal{H}) = \text{e}^{-\mu_{\lambda}}\,{}  \pi(\lambda{}, N_\lambda{}) \prod_{k=1}^{N_{\rm obs}}  N_{\lambda} \int_{\theta} \mathcal{L}^{k}(h^k | \theta) \,{}p(\theta | \lambda )\,{}{\rm d}\theta{}, 
\end{eqnarray}
where $\theta$ are the GW parameters, $N_{\lambda}$ is the number of events predicted by the astrophysical model, $\mu_{\lambda}$ is the predicted number of detections associated with the model and the GW detector, $\pi{}(\lambda{},N_\lambda{})$ is the prior distribution on $\lambda$ and $N_\lambda$, and $\mathcal{L}^{k}(\lbrace h\rbrace^k | \theta)$ is the likelihood of the $k$th detection. The predicted number of detections is given by $\mu{}(\lambda{})=N_\lambda\,{}\beta{}(\lambda{})$, where 
\begin{equation}\label{eq:beta}
\beta{}(\lambda{})=\int_\theta p(\theta{}|\lambda{})\,{}p_{\rm det}(\theta{})\,{}{\rm d}\theta    
\end{equation} 
is the detection efficiency of the model. In eq.~\ref{eq:beta}, $p_{\rm det}(\theta{})$ is the probability of detecting a source with parameters $\theta$ and can be inferred by computing the optimal signal-to-noise ratio and comparing it to a detection threshold, as described, e.g., in \cite{bouffanais2021}. 

The values for the event's log-likelihood are derived from the posterior and prior samples released by the LVC, such that the integral in eq.~\ref{eq:post_hier_model} is approximated with a Monte Carlo approach as 
\begin{equation}\label{eq:approx_integral_likeli}
\mathcal{I}^{k} = \int_{\theta}\mathcal{L}^{k}(h^k | \theta) \,{}p(\theta | \lambda )\,{}{\rm d}\theta{}\sim{}\frac{1}{N_s^k}\,{}\sum_{i=1}^{N_s^k}\frac{p(\theta^k_i | \lambda{})}{\pi^k(\theta_i^k)},
\end{equation}
where $\theta_i^k$ is the $i$th posterior sample for the $k$th detection and $N_s^k$ is the total number of posterior samples for the $k$th detection. Both the model and prior distributions are estimated with Gaussian kernel density estimation. 
We further marginalise eq.~\ref{eq:post_hier_model} over $N_{\lambda}$ using a prior $\pi(N_{\lambda}) \sim 1 / N_{\lambda}$ \citep{fishbach2018}, which yields the following expression
\begin{eqnarray}\label{eq:post_hier_model_marg} 
p(\lambda| \mathcal{H}) \sim \pi(\lambda) \prod_{k=1}^{N_{\rm obs}}  \dfrac{\mathcal{I}^k}{\beta(\lambda)}.
\end{eqnarray}

\begin{figure}
\begin{center}
\includegraphics[width=\columnwidth]{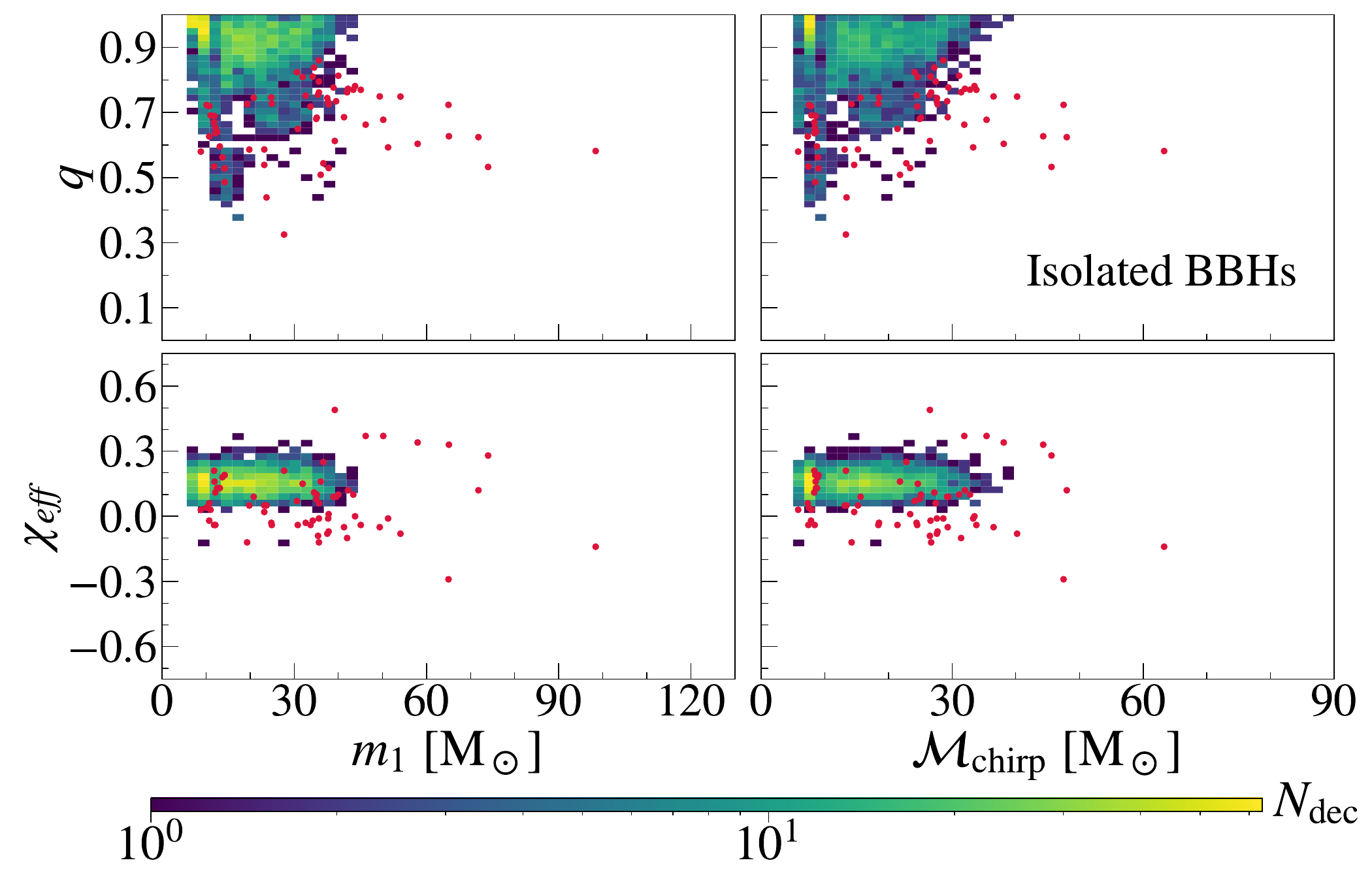}
\caption{Two dimensional histograms of the detectable simulated BBHs in the isolated binary evolution model. The four panels show, from left to right and from top to bottom $q-m_1$, $q-\mathcal{M}_{\rm chirp}$, $\chi_{\rm eff}-m_1$, and $\chi_{\rm eff}-\mathcal{M}_{\rm chirp}$. The red points are the data from \protect\cite{abbottGWTC3population}. We do not add error bars to avoid to make the plot  confusingly crowded.
}\label{fig:histo_iso}
\end{center}
\end{figure}

\begin{figure}
\begin{center}
\includegraphics[width=\columnwidth]{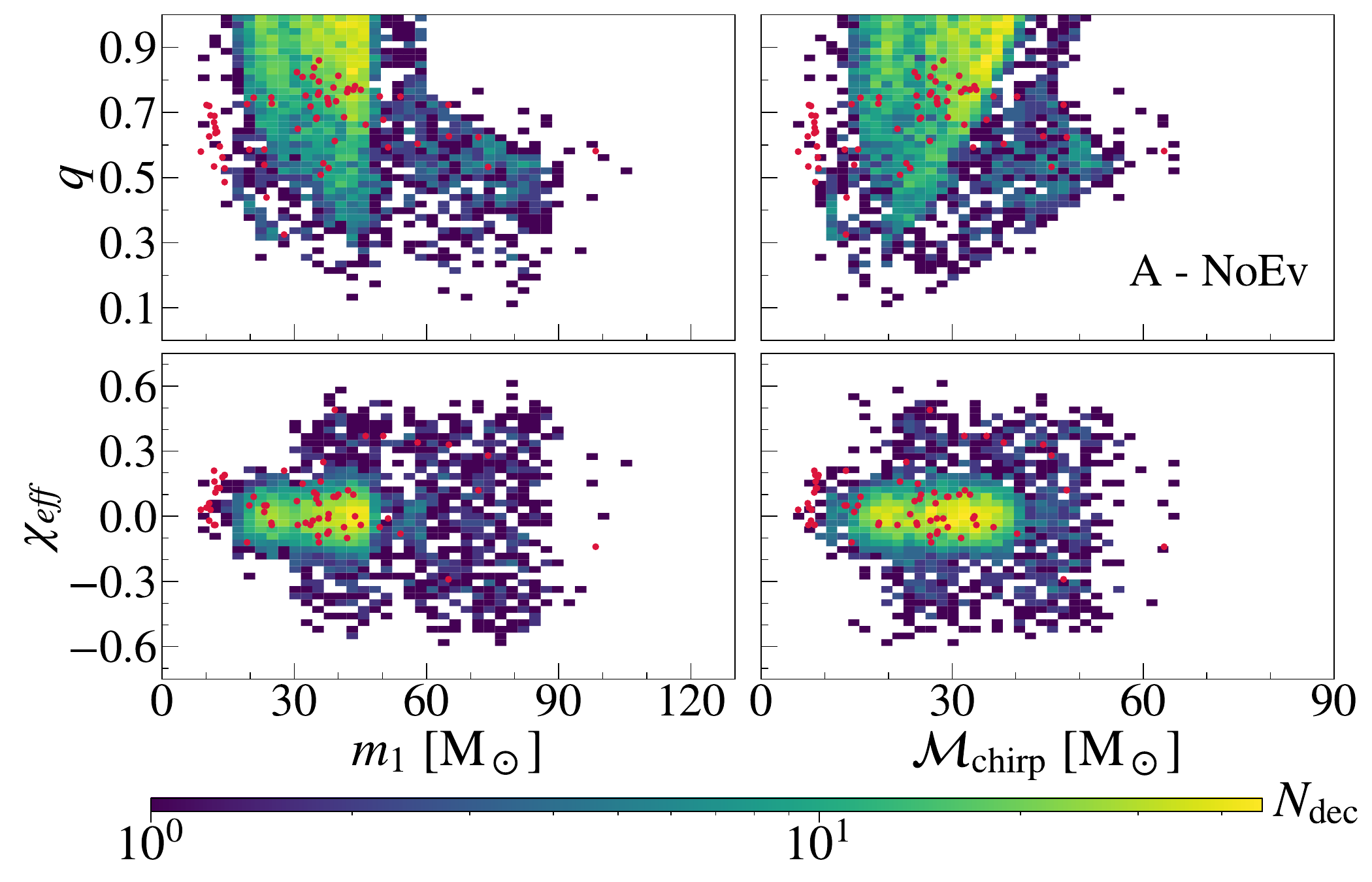}
\caption{Same as Figure~\ref{fig:histo_iso} but for model A NoEv.
}\label{fig:Ong}
\end{center}
\end{figure}

\begin{figure}
\begin{center}
\includegraphics[width=\columnwidth]{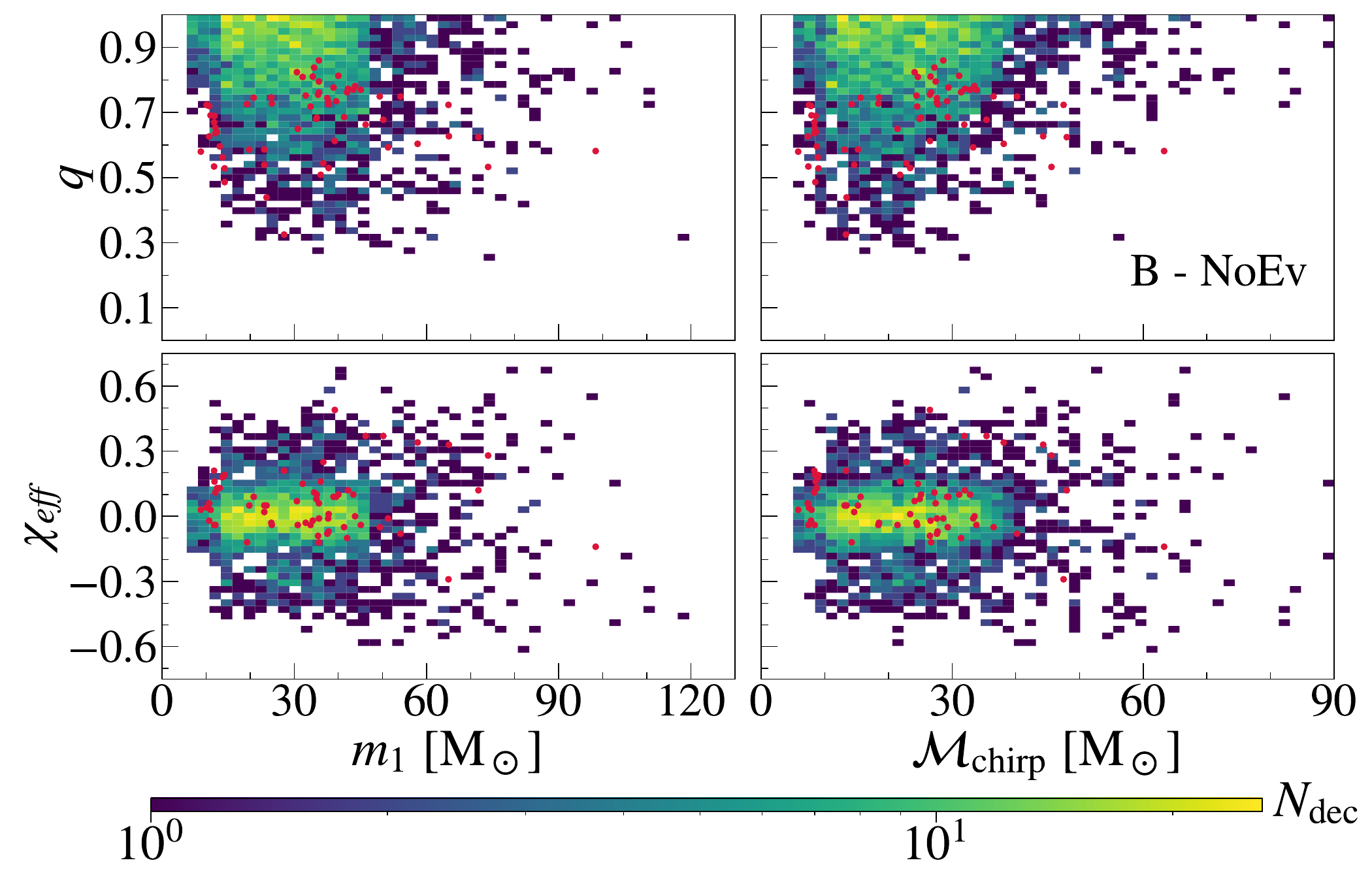}
\caption{Same as Figure~\ref{fig:histo_iso} but for model B NoEv.
}\label{fig:O1g}
\end{center}
\end{figure}

\begin{figure}
\begin{center}
\includegraphics[width=\columnwidth]{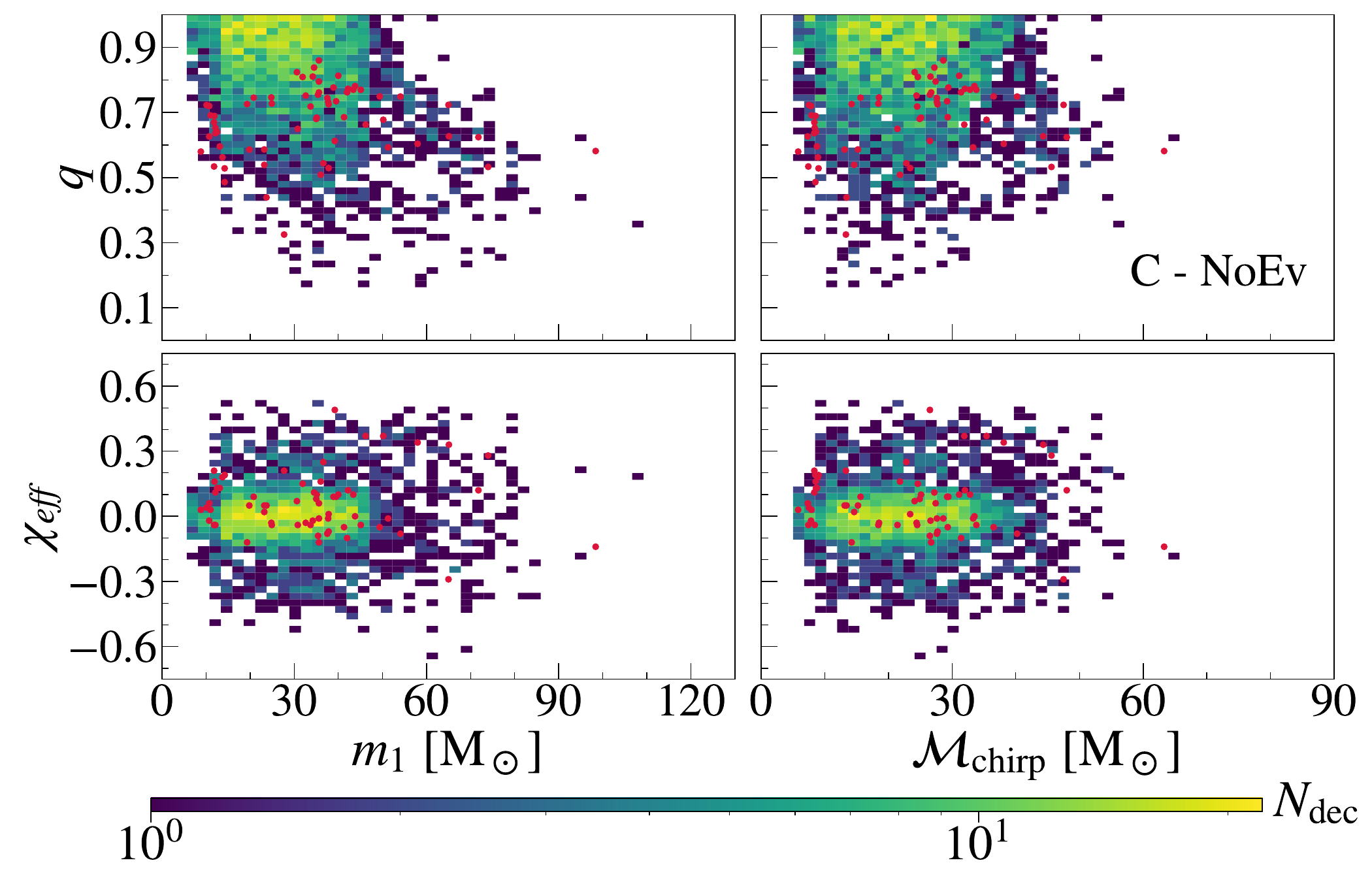}
\caption{Same as Figure~\ref{fig:histo_iso} but for model C NoEv.
}\label{fig:H1g}
\end{center}
\end{figure}

Based on the above calculations, we can thus estimate the mixing fraction as described in eq.~\ref{eq:mixfraction}.
 Figures~\ref{fig:histo_iso}, \ref{fig:Ong}, \ref{fig:O1g} and \ref{fig:H1g} show the behaviour of four selected models (isolated binaries, and models A, B and C in the NoEv case, respectively) in the planes defined by the four parameters $q$, $\chi_{\rm eff}$, $m_1$, and $\mathcal{M}_{\rm chirp}$. In the histograms, we do not show all possible simulated BBHs, but rather only the detectable ones (i.e., with a signal-to-noise ratio\footnote{$\mathrm{SNR}^2 = 4\int^{f_{max}}_{f_{min}}\frac{\tilde{h}^2(f)}{S_{n}(f)}\mathrm{d}f$, where $\tilde{h}$ is the waveform and $S_n$ the power spectral noise density.} above 8 assuming O3b sensitivity). From these Figures, we see that BBHs from isolated binary systems tend to be associated with nearly aligned spins, while our dynamical models have a broader distribution of $\chi_{\rm eff}$, which is favoured by the LVK data. This explains the large values of the mixing fraction $f_{\rm GC}$ we obtain when we include the impact of $\chi_{\rm eff}$ into our hierarchical Bayesian analysis. From these Figures, we also see that isolated BBHs favour a strong peak at $m_1\sim{8-10}$ M$_\odot$, similar to the main peak reported by the LVK collaboration.


\end{appendix}

\end{document}